\documentclass[twocolumn,runningheads]{llncs}
\usepackage{hyperref}
\usepackage{graphicx}
\usepackage{xcolor}
\usepackage[nice]{nicefrac}
\usepackage{amsmath}
\usepackage{amsfonts}
\usepackage{mathtools}
\usepackage{multirow}
\usepackage[textwidth=18.31cm,top=1.5cm]{geometry}

\graphicspath{{./images/}}  
\begin{document}
\title{Medical Image Segmentation on MRI Images with Missing Modalities: A Review}
%
\twocolumn[
  \begin{@twocolumnfalse}
\author{Reza Azad\inst{1} \and
Nika Khosravi\inst{1} \and
Mohammad Dehghanmanshadi\inst{2} \and
Julien Cohen-Adad\inst{3,4,5} \and Dorit Merhof\inst{1,6}}


\institute{Institute of Imaging and Computer Vision,
RWTH Aachen University, Germany\and
School of Automotive Engineering, Iran University of Science and Technology, Tehran, Iran, \and
Functional Neuroimaging Unit, CRIUGM, University of Montreal, Montreal, Canada
\and
NeuroPoly Lab, Institute of Biomedical Engineering, Polytechnique Montreal, Canada
\and
Mila, Quebec AI Institute, Canada\and 
Fraunhofer Institute for Digital Medicine MEVIS, Bremen, Germany\\\email{dorit.merhof@lfb.rwth-aachen.de}}



\maketitle
\begin{abstract}
Dealing with missing modalities in Magnetic Resonance Imaging (MRI) and overcoming their negative repercussions is considered a hurdle in biomedical imaging. The combination of a specified set of modalities, which is selected depending on the scenario and anatomical part being scanned, will provide medical practitioners with full information about the region of interest in the human body, hence the missing MRI sequences should be reimbursed. The compensation of the adverse impact of losing useful information owing to the lack of one or more modalities is a well-known challenge in the field of computer vision, particularly for medical image processing tasks including tumor segmentation, tissue classification, and image generation. Various approaches have been developed over time to mitigate this problem's negative implications and this literature review goes through a significant number of the networks that seek to do so. The approaches reviewed in this work are reviewed in detail, including earlier techniques such as synthesis methods as well as later approaches that deploy deep learning, such as common latent space models, knowledge distillation networks, mutual information maximization, and generative adversarial networks (GANs). This work discusses the most important approaches that have been offered at the time of this writing, examining the novelty, strength, and weakness of each one. Furthermore, the most commonly used MRI datasets are highlighted and described. The main goal of this research is to offer a performance evaluation of missing modality compensating networks, as well as to outline the future strategies for dealing with this issue.

\end{abstract}
  \end{@twocolumnfalse}
]

\section{Introduction}
\label{sec:intro}
Magnetic resonance imaging, widely known as MRI, is one of the most effective techniques used in biomedical imaging for obtaining high contrast images of the soft tissues in human body such as the brain \cite{ouyang2021representation,dinsdale2021learning,feng2021brain}, abdominal organs \cite{conze2021abdominal,azad2021deep,biondetti2021pet,azad2019bi}, legs \cite{reyngoudt2021global,lee2018efficiency}, spine \cite{azad2021stacked}, tissue \cite{bozorgpour2021multi,azad2020attention,feyjie2020semi}, and so on. T1-weighted, contrast enhanced T1 weighted also known as T1c-weighted, T2-weighted, Fluid Attenuation Inversion Recovery (FLAIR), Magnetization Prepared - RApid Gradient Echo (MP-RAGE), and Proton density (PD-weighted) sequences are among the most clinically utilized MRI modalities and our case of interest in this work, each of which reveals distinct characteristics of the human tissue. Each of these sequences provides the medical professional with crucial and complementary information, leading to most accurate diagnoses, consequently followed by the most effective treatment \cite{yao2021mri,pizzi2021mri,yao2021anisamide,bleker2021single,park2021prediction,reza2022contextual}. 

MRI, although being one of the most effective methods for obtaining high-quality images of the human tissue, is prone to artifacts for a variety of causes, which might lead to one or more missing imaging sequence in practice. These artifacts are generally caused by the failure in MRI hardware or the interaction of the patients with imaging devices \cite{graves2013body}. Some examples of these artifacts are the flow of the cerebrospinal fluid in the brain and spinal canal, magnetic susceptibility artifacts and different types of noise \cite{krupa2015artifacts}. \\
Several methods for solving the missing modalities problem in MR images have been presented throughout the years \cite{dalmaz2021resvit,zhang2021modality,hamghalam2021modality,zhou2021feature,zhu2021brain,yu2021mousegan}. Early efforts proposed strategies for synthesizing or impugning absent input data. The synthesis methods usually reconstruct the missing modality images by learning the most important features from the atlas image, and then utilize a classification mechanism to perform voxel-by-voxel intensity prediction \cite{fei2021deep,dalmaz2021resvit}. The latter methods used deep learning and took various approaches to solving the problem. Some of these techniques include translating the modalities to a shared latent subspace \cite{havaei2016hemis}, knowledge distillation \cite{vadacchino2021had,azad2021smu,wang2020multimodal,chen2021learning}, optimizing key feature information across all modalities \cite{zhou2021latent,sylvain2020cross,pan2021collaborative}, and employing conditional generative adversarial networks (cGAN) \cite{sharma2019missing,yu20183d,isola2017image,zhan2021lr}.\\
This review article covers all of the aforementioned main approaches, as well as various unique networks for each direction. In particular, this literature review covers recent work on networks that aim to compensate for missing MRI modalities, and includes all distinct techniques introduced until 2022, which are divided into five groups. As a result, Section 2 includes a taxonomy of the reviewed networks. Later in this section, a brief but relevant description on MRI sequence acquisition, MRI modalities, and probable MRI artifacts is offered. As previously indicated, Section 3 includes a large variety of unique strategies, with a focus on deep learning approaches, which represent the state-of-the-art in the field of semantic segmentation, while dealing with missing MRI modalities. The novelty, strength, weakness, training dataset, network design, and major contributions of each approach are all considered. A comparative overview is also provided at the end of Section 3 to provide the reader with a clearer picture. Section 4 summarizes the most widely used semantic segmentation benchmarks and highlights their key aspects. Section 5 assesses the effectiveness of the proposed methods employing widely used evaluation metrics, which are provided at the start of this section. The obstacles and future solutions for segmentation networks with missing MRI modalities are reviewed in Section 6, and the overall conclusions are presented in Section 7.

\section{Taxonomy}
\begin{figure*}[h]
		\centering
		\includegraphics[width=1\textwidth]{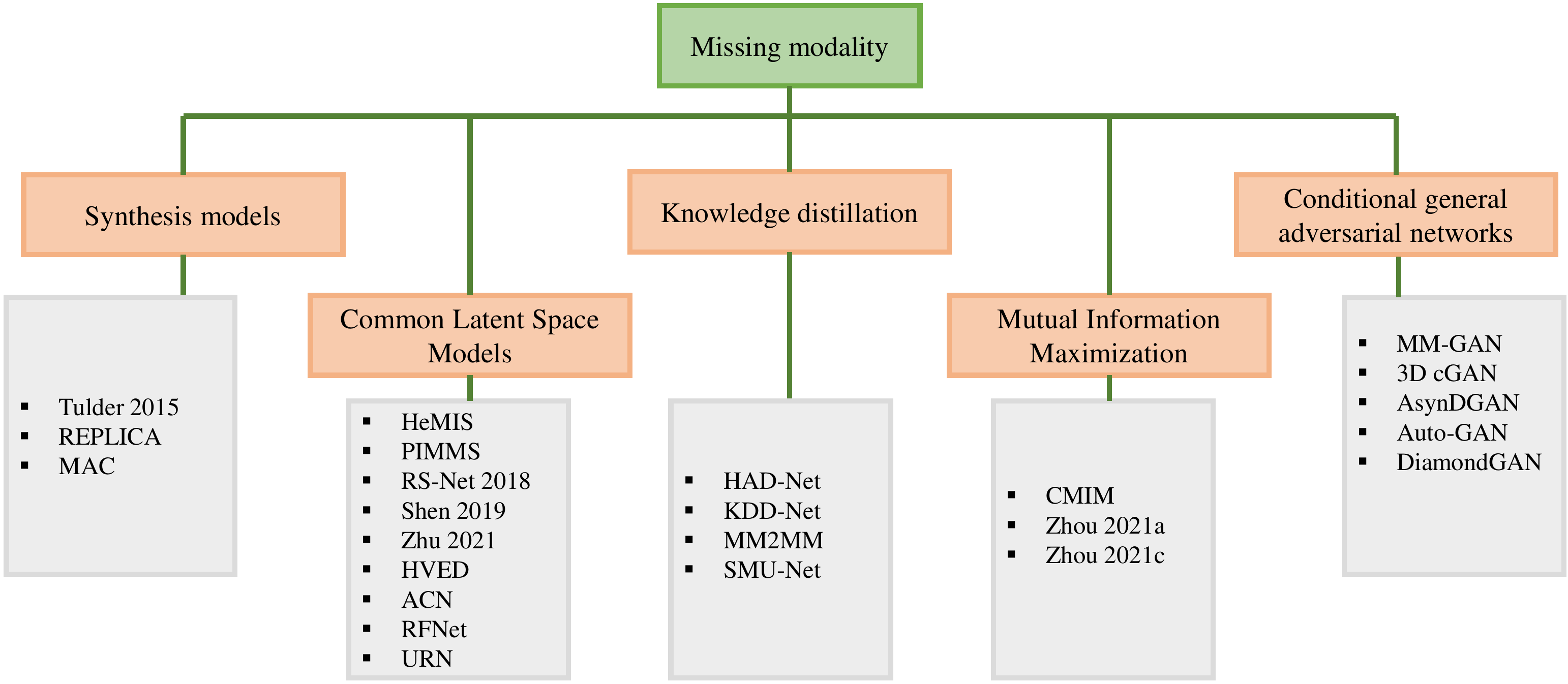}
	\caption{The proposed taxonomy for the reviewed methods on MRI-based semantic segmentation with missing modalities.}
	
	\label{fig:main_approaches}
\end{figure*}

In this section, we present a taxonomy that categorizes different strategies presented in the literature to overcome the problem of missing modality. To this end, a taxonomy of the main approaches performing semantic segmentation on MRI images with missing modalities is depicted in Figure \ref{fig:main_approaches}. We will elaborate on these approaches in Section 3. In the rest of this section, we briefly explain the MRI technique, MRI modalities and some of the artifacts that are commonly seen on MRI data. 

\subsection{Magnetic Resonance Imaging}
The case of interest in this study is MRI, which is frequently utilized for creating high contrast images of soft tissue with high spatial resolution. The inherent difference between pathological and healthy tissue is exposed in MRI scans by adopting the settings and protocols. Other methods such as X-ray and Computed Tomography (CT) expose the body to a certain amount of ionizing radiation. But on the other hand, MRI does not employ X-rays or ionizing radiation, making it unique from the aforementioned techniques. Also the soft tissues of the human brain can be well identified in an MRI scan. As a result, MRI is frequently recognized as the most effective technology for observing and segmenting human tissue, especially the brain.

MRI employs a technique known as Nuclear Magnetic Resonance (NMR), which is based on the magnetisation characteristics of the nuclei of hydrogen atoms ${^1H}$, which are plentiful in the human body, mostly in water but they also exist in fat. A hydrogen atom has a single proton in its nucleus and a single electron orbiting the nucleus. Protons are the main focus of MRI, which takes use of the proton's inherent magnetic properties. Protons spin around their arbitrarily aligned axis, and this spin, also known as magnetic moments, is responsible for proton's magnetic characteristics. An external strong uniform magnetic field that more than 30,000 times stronger than the earth's magnetic field,  is applied to the patient's body during the MRI procedure, which induces the proton spins to align with the field's orientation and consequently, experience precession around the field lines. The Larmor frequency, which is proportional to the intensity of the static magnetic field, is the frequency at which protons precess. Then a radio frequency (RF) current pulse is applied, which results in disturbance of the proton alignment and the coherent precessing of the spins. The induced magnetization is detected by the RF coils, and the spatial information of the image is directly encoded in the Fourier domain thanks to the manipulation of magnetic gradients \cite{webb2017introduction}, \cite{weishaupt2006does}. 

\subsection{MRI Modalities}
There are several MRI sequences, each of which exposes a unique feature of the human tissue. We briefly present the most common sequences:\\
\textbf{T1-weighted}: \\
T1-weighted scans reveal fat within the human body. This means that fatty tissues appear brighter in T1-weighted pictures than other anatomical tissues \cite{Baba.2005a}.\\
\textbf{T2-weighted}: \\
  T2-weighted pictures accentuate water as well as fat, resulting in fat and water tissues appearing bright\cite{Haouimi.2005}.\\
\textbf{Contrast enhanced T1 weighted (T1c-weighted)}: \\
 In contrast enhanced T1-weighted imaging, a gadolinium-based contrast medium is injected into the patient's bloodstream to shorten the T1-relaxation period and better detect lesions with disruption of the blood-brain barrier.\\
 \textbf{Fluid Attenuation Inversion Recovery (FLAIR)}: \\
 The appearance of human body tissue in FLAIR
 scans is similar to that of T2-weighted scans, with the exception that cerebrospinal fluid is depicted dark rather than bright \cite{Baba.2005b}. \\
\textbf{Magnetization Prepared - RApid Gradient Echo (MP-RAGE)}: \\
MP-RAGE scans provide good contrast between white and gray tissues at a relatively short scan time. It is used by a large number of multicenter trials like the Alzheimer's Disease Neuroimaging Initiative (ADNI) \cite{tanner2012fluid}.
\\
\textbf{Proton density (PD-weighted)}: \\
Proton-density weighted sequence or PD-weighted sequence is directly related to the amount of hydrogen nuclei existing in the body region being images. Since fat and fluids contain many protons, tissues appear bright on PD images \cite{yu2015utility}.

\begin{figure}[h]
		\centering
		\includegraphics[width=0.48\textwidth]{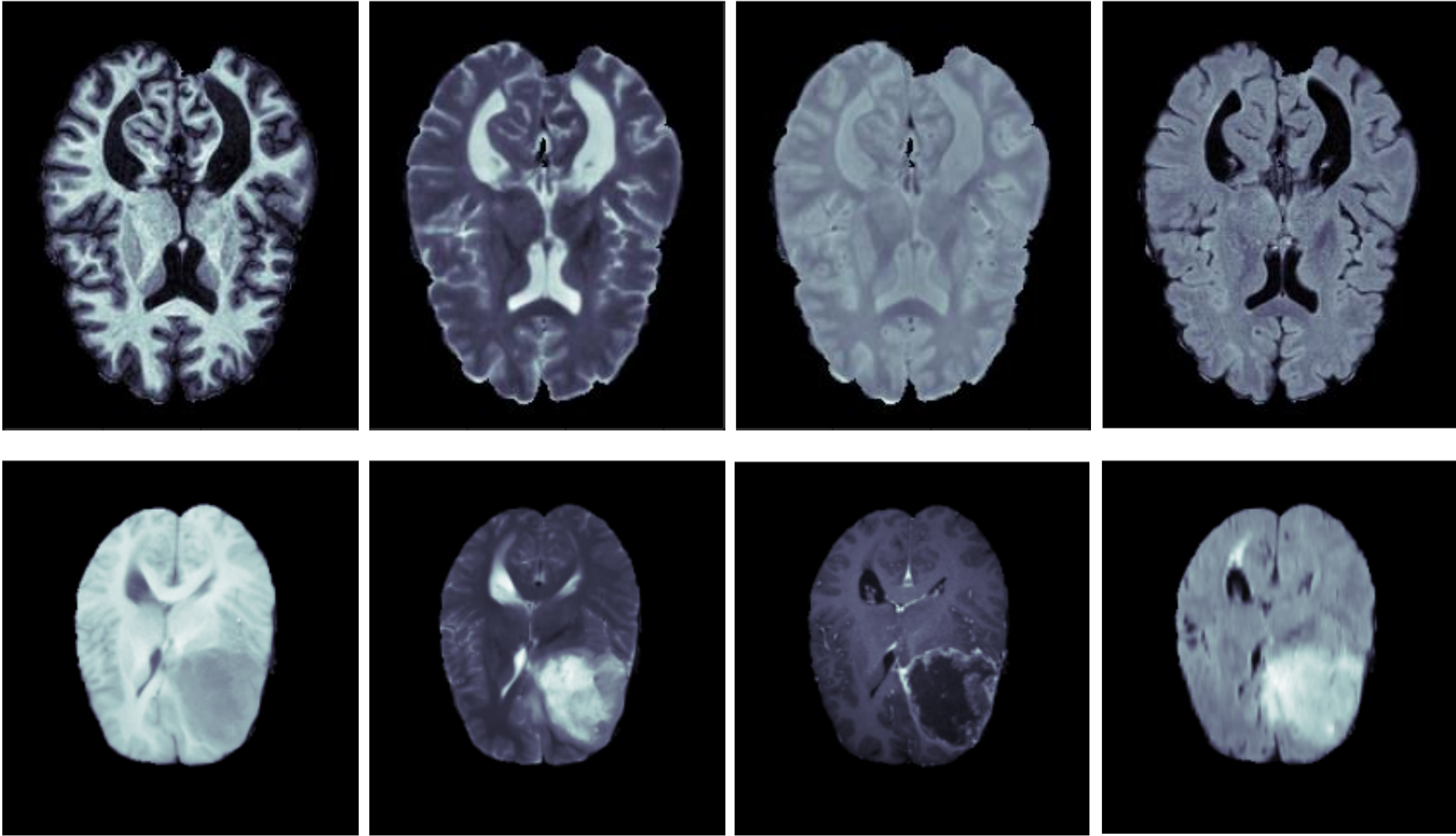}
	\caption{Different MRI sequences are compared. First row \cite{carass2017longitudinal}: MP-RAGE, T2-w, PD-w and FLAIR   are shown from left to right. Second row \cite{menze2014multimodal}: T1, T2, T1c, and FLAIR are shown from left to right. Each MRI modality reveals different characteristics of the soft tissue of the human brain, therefore combining them would provide the clinical professionals with comprehensive information.}
	
	\label{fig:fig2}
\end{figure}

Figure \ref{fig:fig2} depicts a comparison between different types of MRI modalities in two datasets, the Longitudinal multiple sclerosis lesion dataset \cite{carass2017longitudinal} and the BraTS dataset \cite{menze2014multimodal}.
\subsection{MRI Artifacts}
The MRI artifacts may be divided into three categories. The first category includes artifacts induced by movement, such as respiratory motion, blood flow motion and flow of the cerebrospinal fluid in the brain and spinal canal. The second type of artifacts might arise if the measuring technique or settings are not chosen precisely. Aliasing, chemical shift, phase cancellation, coherence and magnetic susceptibility artifacts are examples of this category. Finally, exogenous sources like magnetic field distortions, the hardware itself and noise may cause the third type of artifacts \cite{brown2011mri}.
It is worth noting that some of the aforementioned factors have a substantial impact on MRI scans and could result in missing of one or more modalities, which is known as a common challenge in MRI. Please refer to \cite{krupa2015artifacts} for more details on the MRI artifacts. 

\begin{figure}[h]
	\centering
	\includegraphics[width=0.5\textwidth]{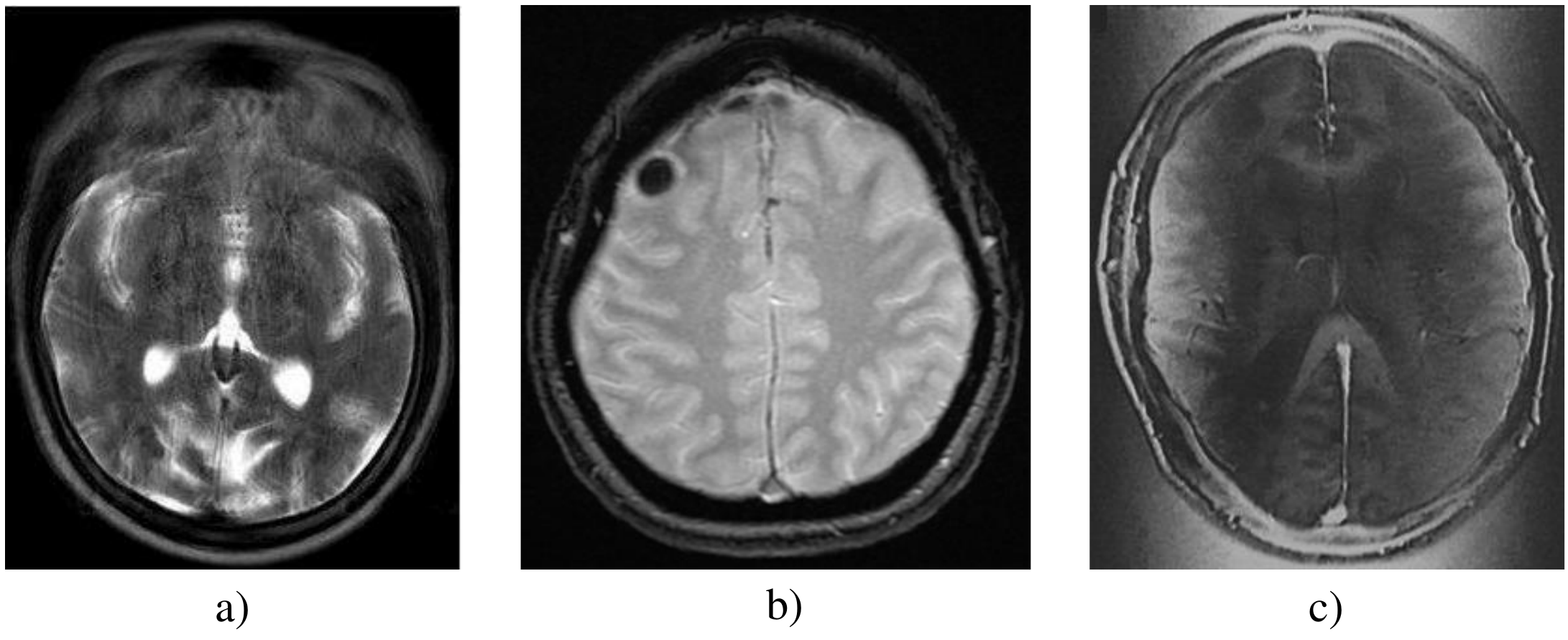}
	\caption{The most frequent MRI artifacts. a) caused by the patient's movements in a MRI scan of the brain \cite{article.motion} b) Magnetic susceptibility artifacts in a MRI scan \cite{web:lang:stats}  c) RF overflow artifact, which occurs due to MRI hardware \cite{web:lang:stats2}}
	\label{fig:artifacts}
\end{figure}

Figure \ref{fig:artifacts} depicts three of the most frequent MRI artifacts and how they impair MRI scans by reducing the visibility and detectability of the region of interest in the human brain in the obtained scans. 

\subsection{Definition and Clinical Effect of Missing Modalities}
Earlier we introduced the MRI modalities and different causes might result in missing one or more MRI modalities. We refer to the missing modality as a problem where one or more modalities are missing in the inference time while the training time has access to the full modality (e.g. T1w, Tw2, T1c and Flair) dataset. In our definition, we follow the literate work and consider each MRI sequence (or contrast) as a unique modality and discuss approaches are proposed to address the missing modality in the segmentation task. To point out the importance of the missing modality in medical treatment we refer to the \cite{brady2017error}. According to \cite{brady2017error} annotating the object of interest (e.g. brain tumour) in MRI images always come with uncertainty and mistakes, thus, both humans and algorithms may contribute to the deteriorating of the radiologist's performance. Statistical information shows that human-level mistakes during the annotating process can deteriorate the radiologist's work up to $21\%$ in MRI images \cite{brady2017error}. The process may even get worsen if one or more modalities are missing since each modality contains specific information that might not be completely recovered using the remaining modalities. Therefore, an automatic algorithm that can compensate for the missing modality plays a significant role in clinical application and our case study in this review paper.

\section{Missing Modality Compensating Networks}
As illustrated in Figure 1, the solutions utilized to overcome the missing modality problem can be divided into five main categories:
\begin{itemize}
    \item Synthesis models
    \item Shared latent space
    \item Knowledge distillation networks
    \item Mutual information maximization
    \item Conditional general adversarial networks (cGANs)
\end{itemize}

In the following parts of Section 3, each of the five main categories will be discussed extensively, with numerous prominent networks as examples introduced and analyzed for each.

\subsection{Synthesis Models}
Early attempts to overcome the problem of missing modalities included synthesizing the missing modality. \cite{van2015does} conducted experiments on two different classifiers: 1) Support vector machines (SVMs) 2) Random forests (RFs), using two models, neural networks and restricted Boltzmann machines (RBMs), that synthesize the missing modality. 
The neural network used in \cite{van2015does} is a simple feed-forward network with just three layers that is able to predict a 3D patch. 
The second model used in \cite{van2015does} is an RBM. RBMs are restricted Blotzmann machines, and they also be regarded as stochastic neural networks that can learn critical characteristics of a probability distribution using relevant information from an unidentified probability distribution \cite{fischer2012introduction}.

As shown by \cite{van2015does}, inferring missing data at test time using a synthesis method, which is more adaptive than the classifier, may improve multi-modal image segmentation by supplying data transformations for the classifier and also enlarging the training set. Through the use of synthesized data, random forest, a basic classifier presented in \cite{van2015does},  has exhibited improvements in segmentation outcomes.

\begin{figure}[h]
		\centering
		\includegraphics[width=0.5\textwidth]{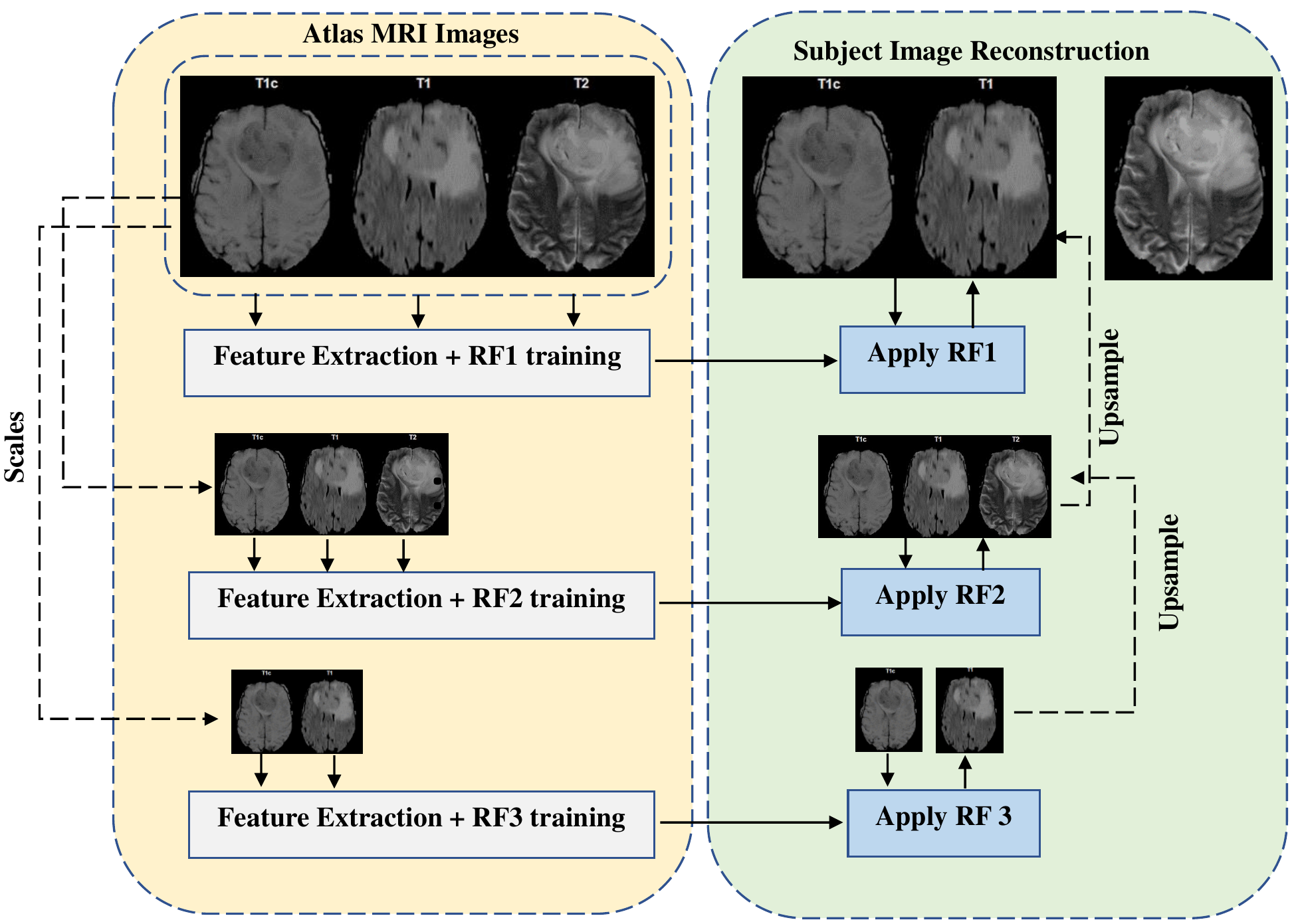}
	\caption{The REPLICA algorithm \cite{jog2017random} synthetizes T2w images from T1w images by first extracting a set of high-level features and then training a random
forest at different scales. By matching a feature set in various scales it synthesis the T2w image.}
	\label{fig:fig9}
\end{figure}

Figure \ref{fig:fig9} depicts a model called REPLICA, which stands for Regression Ensembles with Patch Learning for Image Contrast Agreement and is introduced in \cite{jog2017random}. REPLICA is a supervised random forest non-linear regression approach for synthesising the missing modalities that is capable of synthesizing T2 and FLAIR, which was previously thought to be a hurdle. REPLICA is structured to forecast tissue contrasts based on inputs with the same tissue contrast as the MRI image to be produced.

The REPLICA architecture shown in Figure \ref{fig:fig9} predicts  T2-weighted images from T1-w images. The training process is depicted on the left side, which is taking place on three different scales. Firstly, the most important features from the atlas image set are extracted in each scale, and the random forest is then trained to predict voxel-by-voxel intensity.
After the training process is completed, REPLICA aims to synthesize a missing modality, with the following description: starting from the coarsest scale, the trained random forest in scale 3, is applied to the features extracted in the same scale, in order to synthesize the target MRI scan. With this scale synthesizing at the lowest resolution, the features extracted from this scale are then up-sampled to the next level. The procedure proceeds to scale 1, which is the finest scale, with high resolution features, and then the trained random forest RF1 generates the final synthetic image.

In \cite{hofmann2008mri}, a model is described that generates pseudo-CT images from brain MR images, which are then utilized for tissue photon attenuation correction (AC). This MRI-based AC for PET/MRI scanners incorporates two approaches for producing the attenuation map, namely a pattern recognition method using a Gaussian process regression that leverages local information and an atlas registration, which is essentially an atlas database comprising brain MR and CT scans from 17 patients that utilizes global information.

\cite{van2015does} has evidenced through experiments that in certain scenarios, substituting the missing modality with a synthesized or imputed sequence would not aggravate the findings, but may result in no improvement. For example, it was demonstrated in \cite{van2015does} that using a sequence generated by a three-layered feed forward neural network produced the same results as simply replacing it with zeros or performing the segmentation without it.

\subsection{Common Latent Space Models}

\begin{figure}[h]
		\centering
		\includegraphics[width=0.5\textwidth]{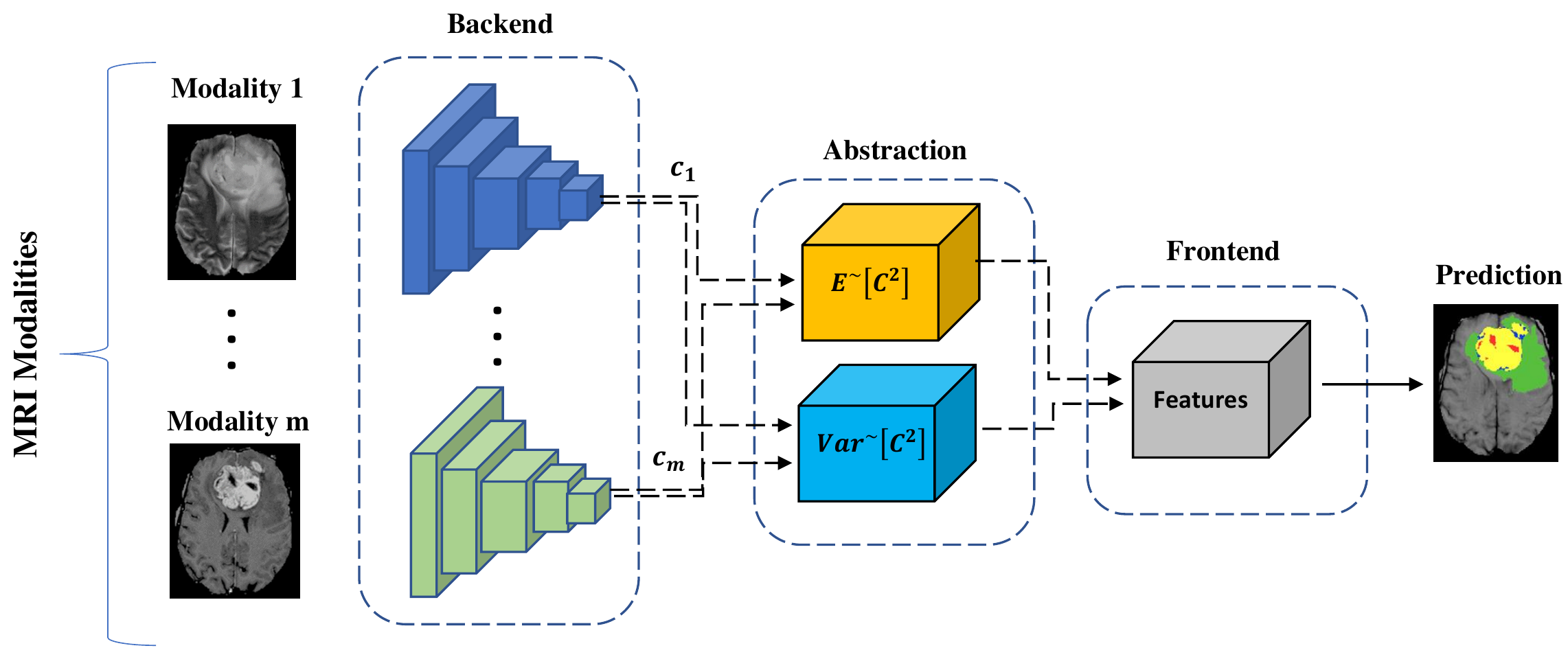}
	\caption{The HeMIS architecture \cite{havaei2016hemis} applies a series of three connected blocks: a Back-end block to encode each modality into a latent space and learn modality-specific features, an Abstraction block to extract statistical features (first and second-order moments) and finally a Front-end block to generate the segmentation map based on the learned representation.}
	\label{fig:fig10}
\end{figure}

Adopting deep learning for biomedical image segmentation  was one of the significant steps for finding a viable strategy for dealing with missing modality issues in MR images.
The objective of early deep learning methods for the missing modalities issue was to translate modalities to a shared subspace and create a shared latent vector. Hetero-Modal Image Segmentation (HeMIS) \cite{havaei2016hemis} is a well-known example that utilizes this concept. HeMIS consists of three main layers as it is also shown in Figure \ref{fig:fig10}: back-end layer, abstraction layer and front-end layer. Each modality will be directed into a specific set of convolutional layers in the network's back-end layer, which will subsequently translate each modality into a common representation of all modalities. Arithmetic operations like mean and variance will be computed in the abstraction layer. The mean and variance will then be combined and supplied into the front-end layer, which will provide the segmentation outputs. 

\begin{figure}[ht]
		\centering
		\includegraphics[width=0.5\textwidth]{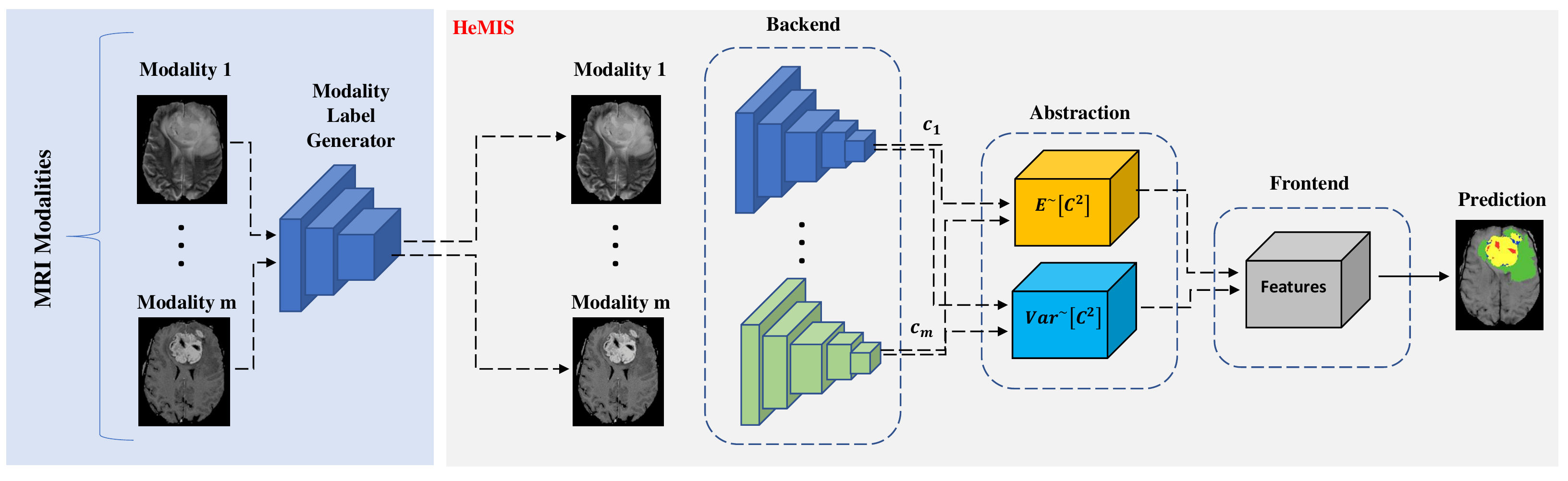}
	\caption{An illustration of the PIMSS method \cite{varsavsky2018pimms}, which is designed to tackle the problem of missing modality and labels. At first, it applies a $f_{mod}$ function to generate a new representation for each modality using a joint representation and then it deploys a HeMIS approach to perform semantic segmentation.}
	\label{fig:fig11}
\end{figure}

Although establishing a common latent embedding for all available modalities is one of HeMIS' major goals, computing the mean and variance alone will not always suffice.
Besides that HeMIS can only function properly in the absence of modalities if each modality input in the test set is labeled. 
The authors of \cite{varsavsky2018pimms} were inspired by the aforementioned HeMIS problem to create a network that, in addition to missing modalities, tackles the issue of missing modality labels. Figure \ref{fig:fig11} shows a HeMIS modification called Permutation Invariant Multi-Modal Segmentation (PIMMS), which can perform segmentation tasks without using modality labels. PIMMS uses a classifier to build a distribution across modalities for the available inputs, then awards a score and labels each unlabeled input data. The inputs are then further adjusted by applying two different types of attention: soft and hard attention. The adjusted inputs are subsequently supplied into the second part of the network, which is a HeMIS model.

\begin{figure}[h]
		\centering
		\includegraphics[width=0.49\textwidth]{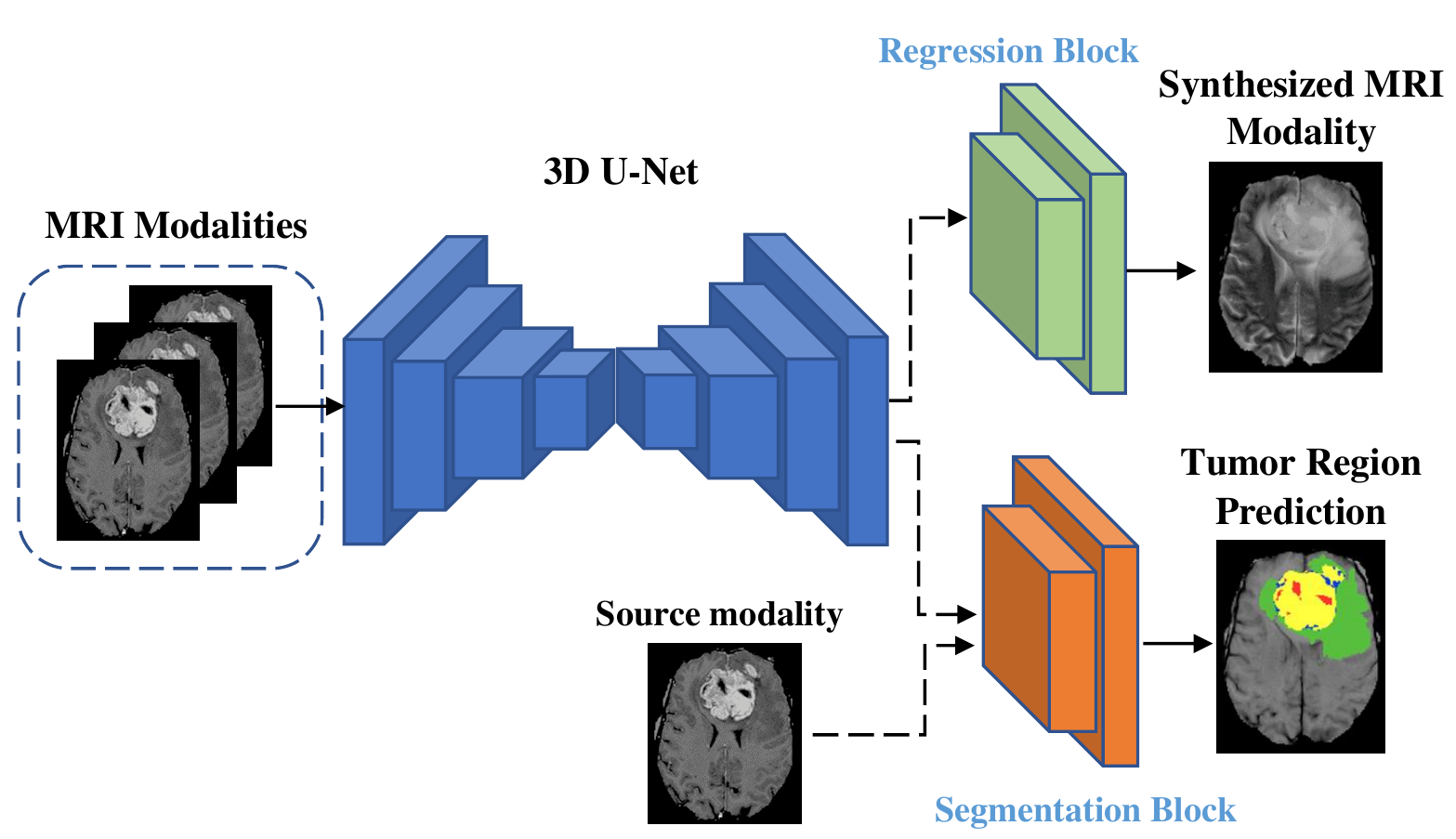}
	\caption{The RS-Net architecture \cite{mehta2018rs} performs both semantic segmentation and MRI modality synthesizing by deploying a 3D U-Net model to represent the input modalities in high-level representation space. It then performs the segmentation and synthesizing by utilizing convolutional and regression heads.}
	\label{fig:fig12}
\end{figure}

The Regression-Segmentation 3D CNN (RS-Net) \cite{mehta2018rs} is another method that builds a common representation of all modalities and synthesizes the missing modality. Three major blocks make up RS-Net. The first block is a 3D U-Net, which is very similar to the one presented in \cite{cciccek20163d}. As illustrated in Figure \ref{fig:fig12}, the U-Net will be fed all of the present volumetric data, resulting in an intermediate latent representation of the data. The second block, the regression convolution block, synthesizes the missing modality through using latent representation and also one of the existing volumetric data as input. The third block is the segmentation convolution block, which takes the produced latent representation as input and has multiple segmentation classes as outputs, each indicating a tumor subtype. The main shortcoming of RS-Net is that it results in errors while attempting to synthesize the T1c modality.

\begin{figure}[h!]
		\centering
		\includegraphics[width=0.5\textwidth]{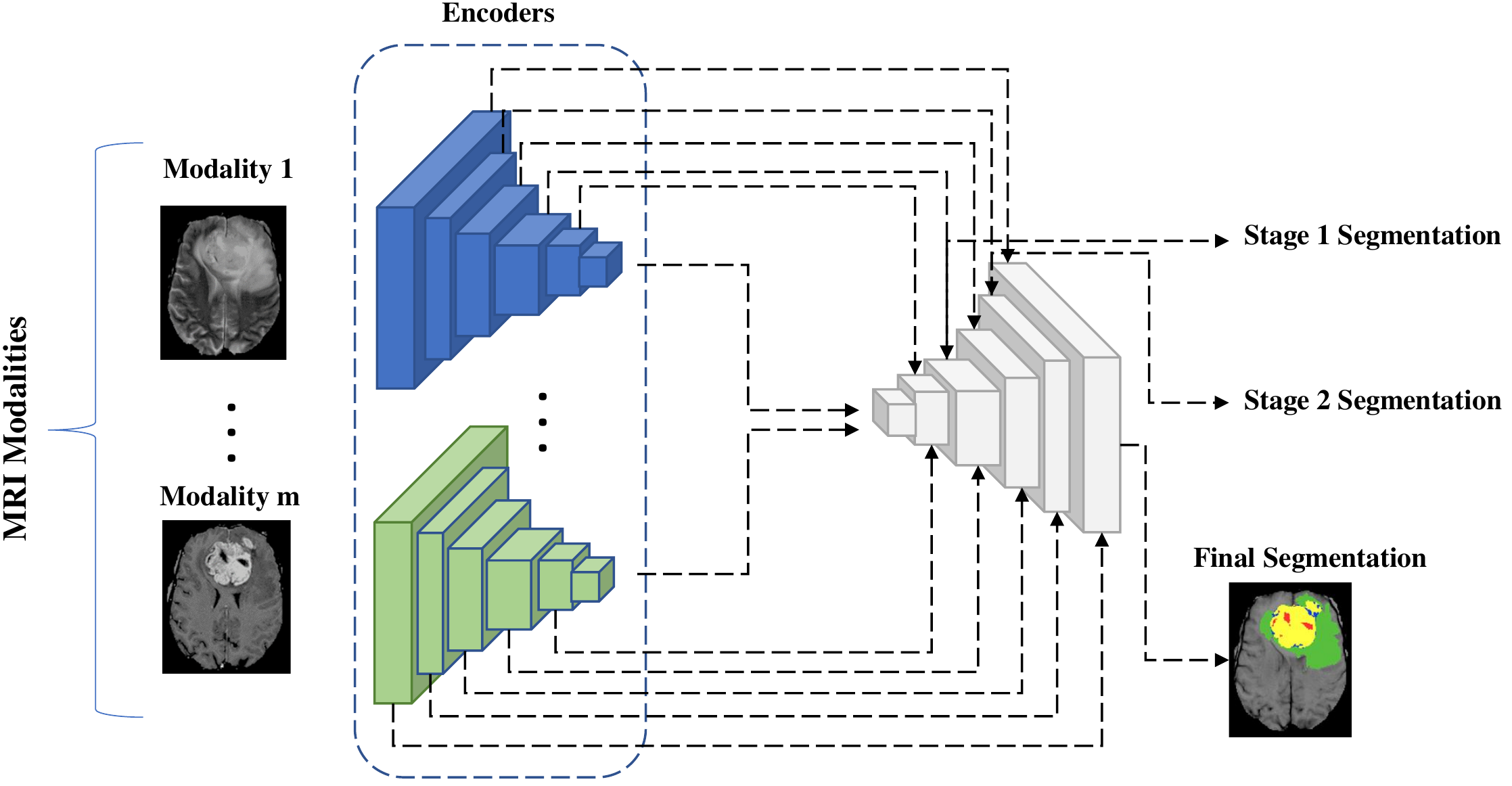}
	\caption{The architecture of the segmentation network proposed in \cite{shen2019brain}, which utilizes a four parallel encoding path.}
	\label{fig:fig13}
\end{figure}

Another method that is showcasing a U-Net-based structure is introduced in \cite{shen2019brain}. The network depicted in Figure \ref{fig:fig13} has four separate encoding paths for obtaining initial feature maps for each MRI modality.
Then, the final segmentation map of the missing modality is generated by combining the initial feature maps and fusing them with feature maps retrieved along the decoding path at multiple resolutions.

\begin{figure}[h]
		\centering
		\includegraphics[width=0.5\textwidth]{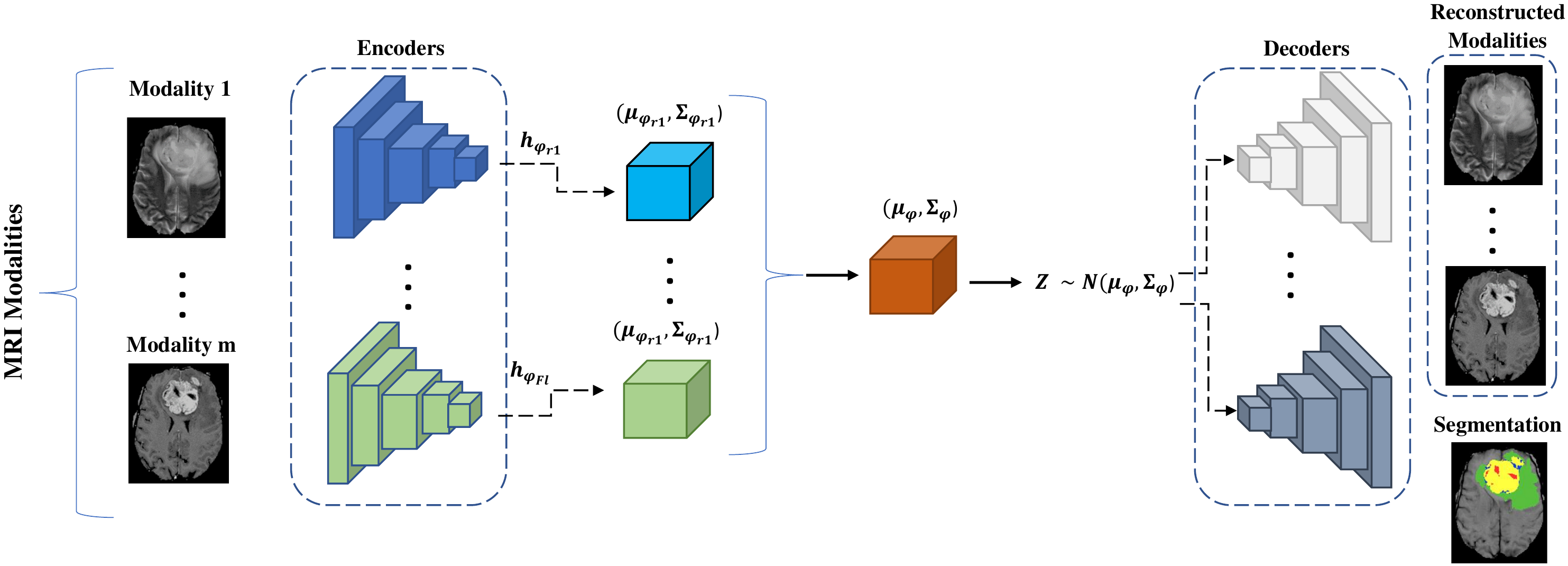}
	\caption{ Learning the common space using variational auto encoder model in MVAE structure \cite{dorent2019hetero}.}
	\label{fig:fig14}
\end{figure}

\begin{figure}[h]
		\centering
		\includegraphics[width=0.5\textwidth]{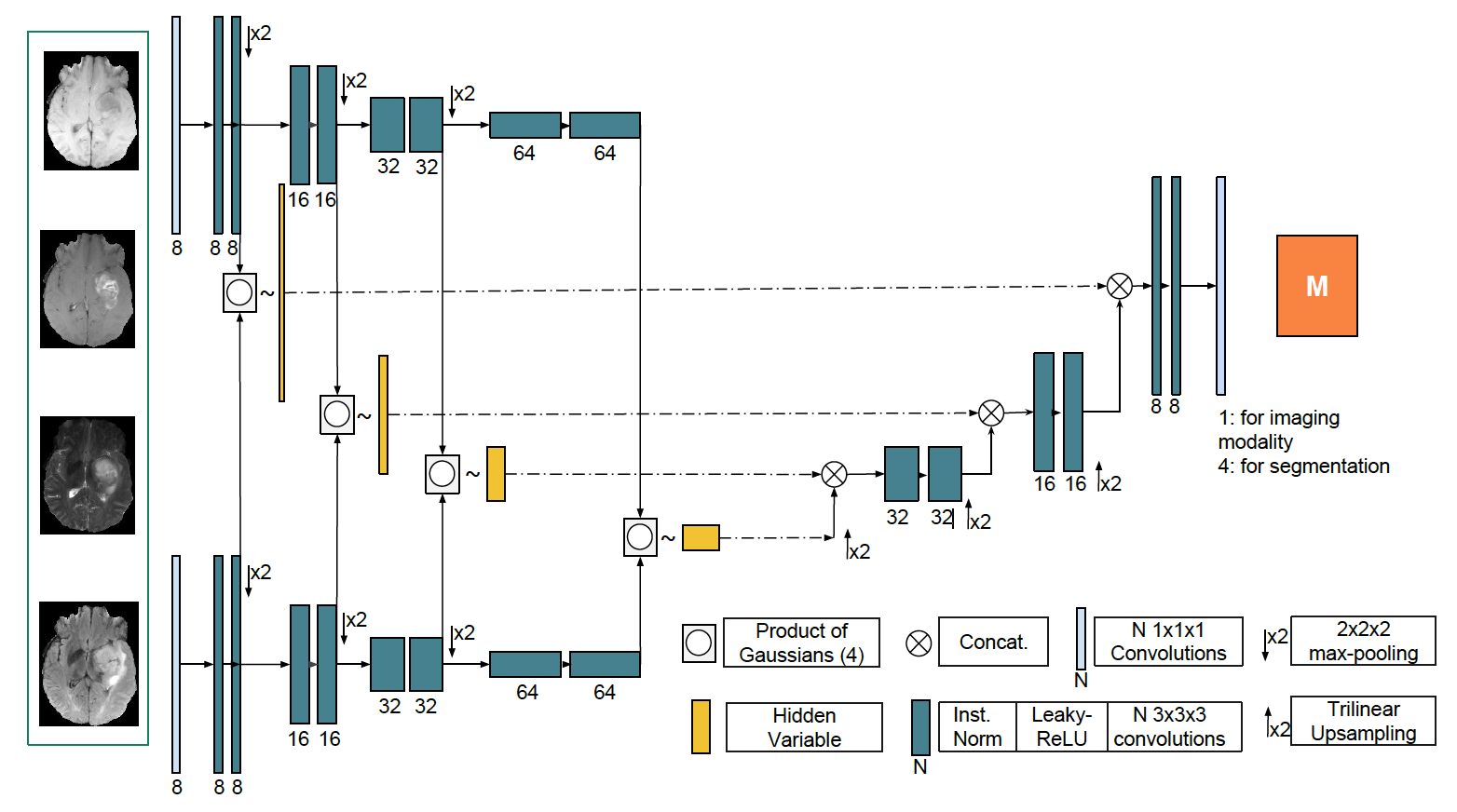}
	\caption{ Multi-level common feature learning in HVED architecture \cite{dorent2019hetero} to reconstruct the missing information.}
	\label{fig:fig15}
\end{figure}

Subsequent attempts to create a shared latent space representation resulted in the development of the Hetero-Modal Variational Encoder-Decoder (HVED), which is a combination of the 3D U-Net and the Multi-Modal Variational Auto-Encoder (MVAE) \cite{dorent2019hetero}. The proposed MVAE architecture, which is depicted in Figure \ref{fig:fig14}, contains four encoders, each of which separately computes the variational parameters, or more specifically, mean and variance of each inference network, which are describing as a gaussian distribution after being merged \cite{wu2018multimodal} and forms a common subspace. 
Following that, five decoders decode a randomly chosen latent variable from the common subspace, see Figure \ref{fig:fig15}. The first four decoders generate the desired modalities, while the fifth generates the segmentation map. In spite of the fact that HVED outflanks HeMIS and U-HeMIS (a HeMIS extension) it produces relatively inadequate results when more than one modality is lacking \cite{wang2021acn}. 

\begin{figure}[h]
		\centering
		\includegraphics[width=0.5\textwidth]{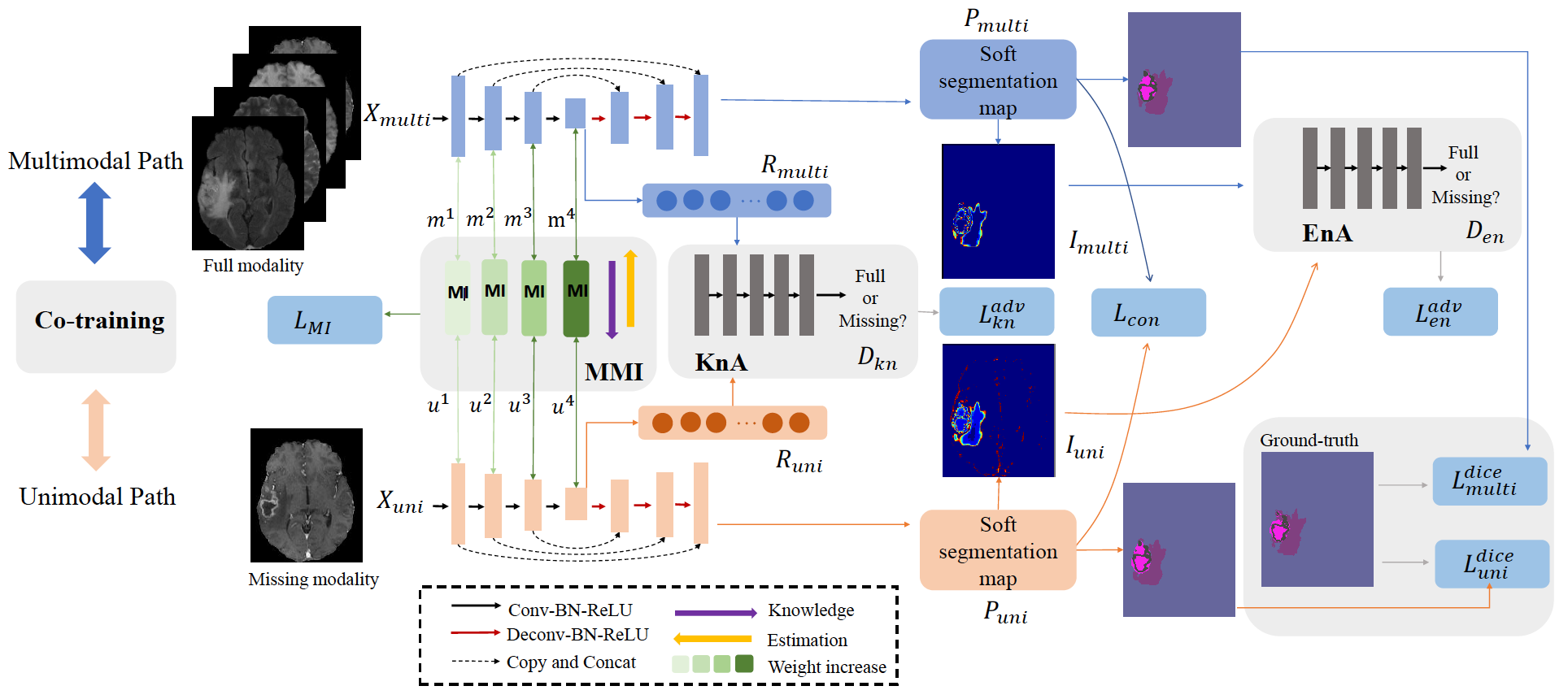}
	\caption{ An illustration of the ACN architecture. The ACN method learns a common latent representation by deploying multimodal and unimodal paths with a co-training approach. To encourage feature matching in different levels of the representation space, it utilizes an entropy adversarial learning module (EnA); a knowledge adversarial learning module (KnA) and a modality-mutual information transfer module (MMI) \cite{wang2021acn}.}
	\label{fig:fig16}
\end{figure}

Calculating the first and second moments is not the only technique to arrive at a shared latent representation. This aim might likewise be achieved using adversarial methods. In \cite{wang2021acn} a model referred to as Adversarial Co-Training Network (ACN) is proposed. The ACN architecture depicts a multimodal path with complete modalities and a unimodal path with the incomplete modality as inputs, see Figure \ref{fig:fig16}. Each path is trained individually and passes through a U-Net on its own. Then the  joint learning process is occurred by embedding an entropy adversarial learning module (EnA), a knowledge adversarial learning module (KnA) and a modality-mutual information knowledge transfer module (MMI) into the network’s architecture. \\
The Segmenatation Maps created by each path are fed into the EnA module, which is located at the end of networks. At each training epoch, the EnA will act as an adversarial discriminator, assisting the two networks in producing increasingly similar segmentation maps. 
The adversarial loss is calculated by the KnA module, which, like EnA, assists the two networks in having more similar outputs. 
The MMI module's role is to compute the Mean Squared Error (MSE) and prevent feature information loss for the path with multi-modal network.

\begin{figure}[h]
		\centering
		\includegraphics[width=0.5\textwidth]{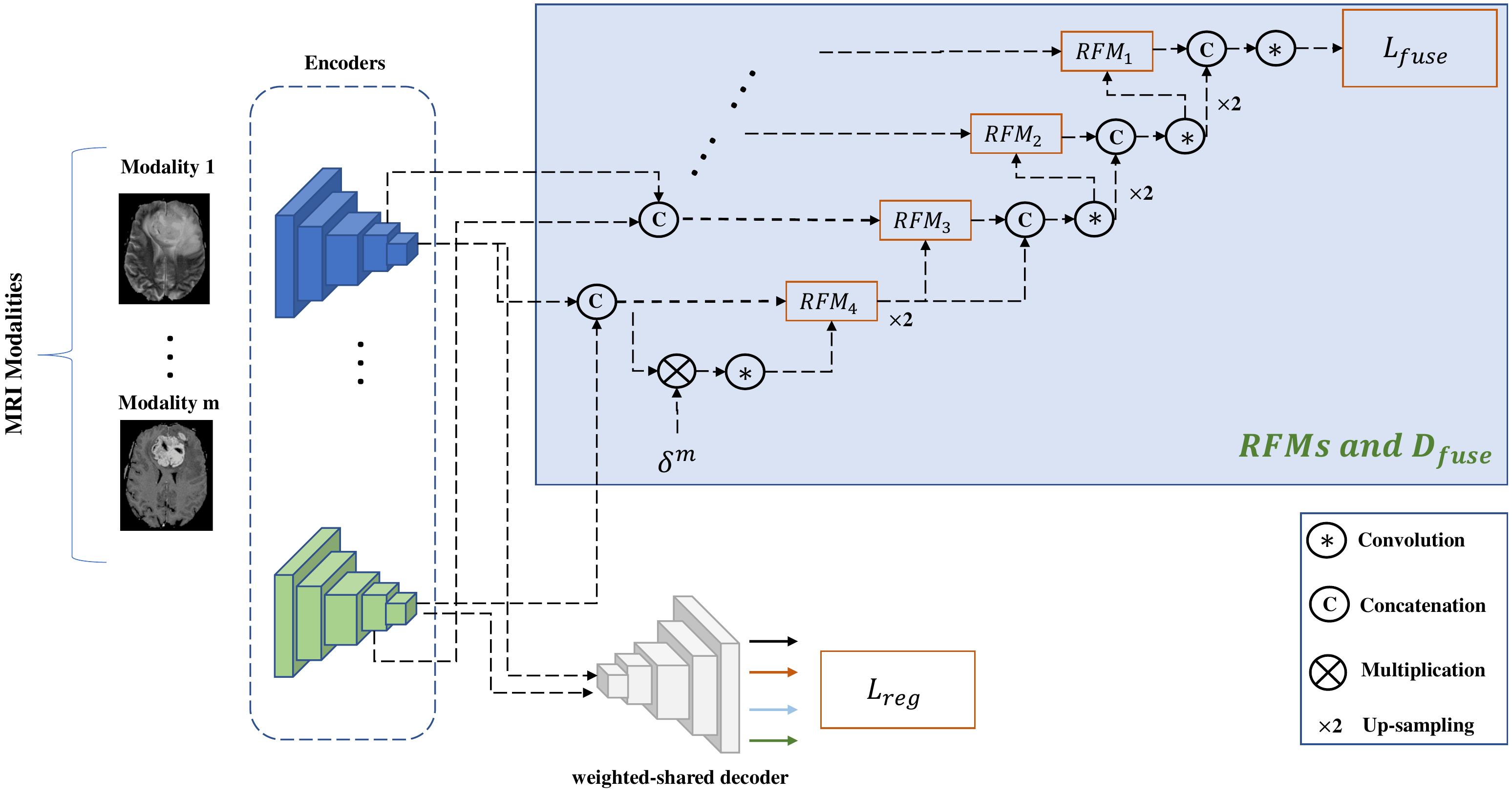}
	\caption{The architecture of the RFNet. Four parallel encoders, i.e., EFlair, ET1c, ET1 and ET2, are utilized to perform the feature encoding operation. The $D_{sep}$ perform the segmentation task while the $D_{fuse}$
is designed to capture the shared features among modalities \cite{ding2021rfnet}.}
	\label{fig:RFNet}
\end{figure}

The authors of \cite{ding2021rfnet} present RFNet, a feature fusion network. RFNet includes four encoders, each of which extracts features from a single modality, as seen in Figure \ref{fig:RFNet}. Then, in order to build a shared representation, a decoder that also shares the weights for the four modalities segments each modality individually. The retrieved features are then fused at different levels using a Region-aware Fusion Model (RFM), and the produced fused representation is then segmented.

\begin{figure}[h]
		\centering
		\includegraphics[width=0.5\textwidth]{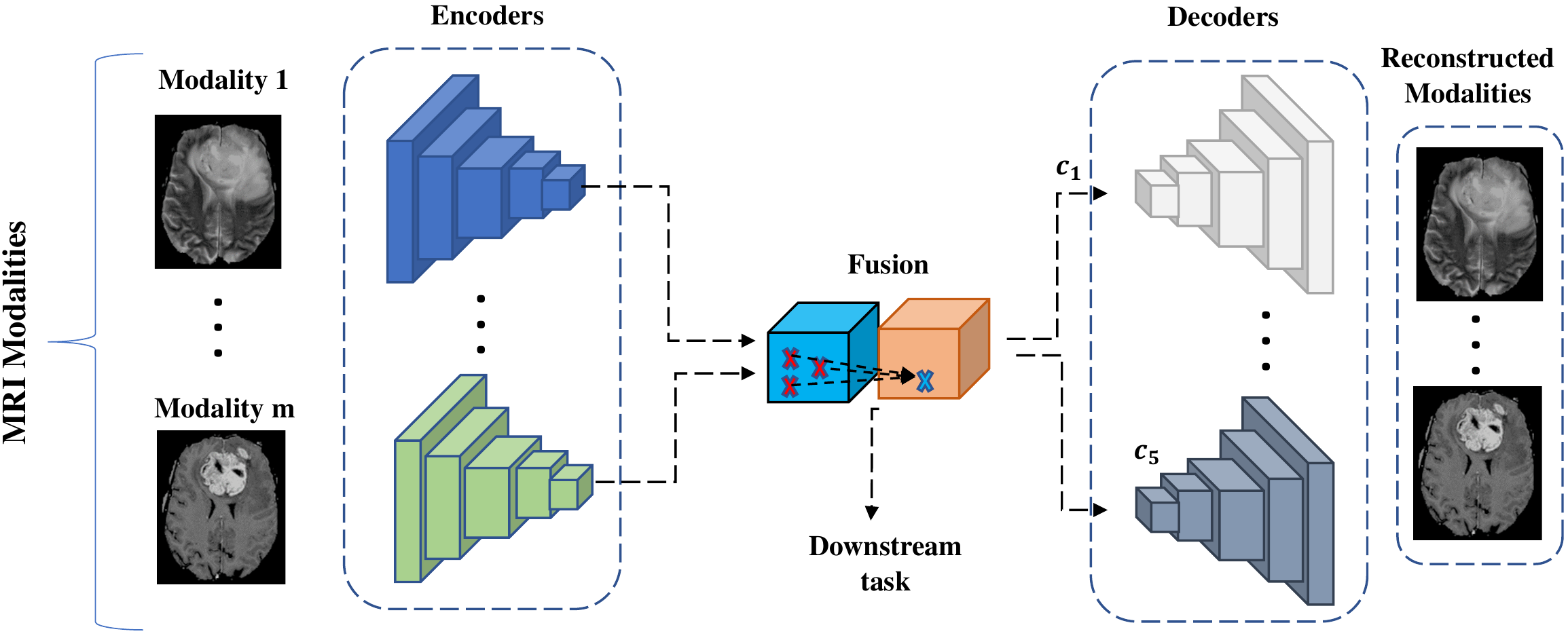}
	\caption{URN architecture \cite{lau2019unified}.}
	\label{fig:Unified}
\end{figure}

As seen in Figure  \ref{fig:Unified}, the method presented in \cite{lau2019unified} is a relatively simple feature fusion method. The Unified Representation Network (URN) uses a U-Net to encode each modality independently, then uses a fusion module to combine the encoder's output. Following that, the newly formed unified representation will be utilized to reconstruct and synthesize the missing modality.

\subsection{Knowledge Distillation Networks}
The authors of \cite{bucilua2006model} presented ``Model Compression'' in 2006 as a novel strategy that allows more simple and faster models to actually learn from a larger, more sophisticated, and higher performing one. 
The authors of \cite{hinton2015distilling} were later inspired by this approach and they recommended to train a more intricate and larger model with also more accurate results, and then transfer the information gained from this model to a smaller model with the aim of enhancing its performance. The proposed approach was named ``Knowledge Distillation'', and it has subsequently been used in a variety of networks. This section outlines the methodology of two networks that employ knowledge distillation.

Hierarchical Adversarial Knowledge Distillation Network (HAD-Net) is a network proposed by  the authors of \cite{vadacchino2021had} that uses the benefits of hierarchical adversarial training.


HAD-Net is comprised of three subnetworks: the teacher network, the student network, and the hierarchical discriminator (HD).
The teacher network is a 3D U-net \cite{isensee2018no} and that has been trained with the full modalities. The same 3D U-Net is used to form the student network as it is for the teacher network, however the T1c modality is not used during inference. At varying resolutions, the fully convolutional HD is in charge of distilling information from the teacher network onto the student network. Hereafter a Mean Squared Error (MSE) adversarial loss is computed between the labels generated by the teacher network and the student network.

\begin{figure}[h]
		\centering
		\includegraphics[width=0.5\textwidth]{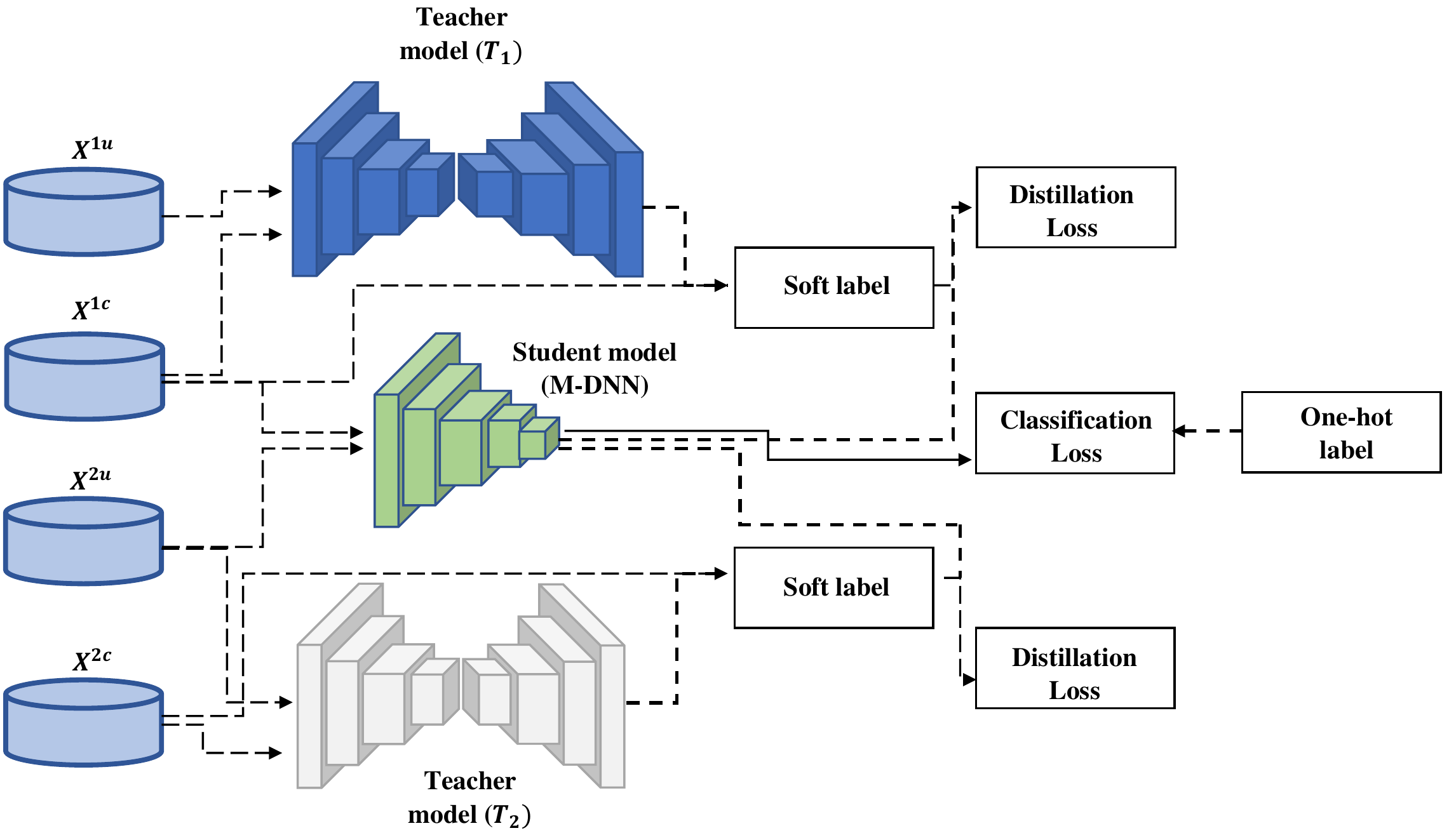}
	\caption{ The knowledge distillation pipeline proposed in \cite{wang2020multimodal}. At first, it trains the teacher model using both full and missing modality samples. Then using the output of the teacher model as a soft label along with the one-hot labels trains the student model.}
	\label{fig:KDD}
\end{figure}

The network suggested in \cite{wang2020multimodal} will be explained in order to further review the knowledge distillation models. The authors of \cite{wang2020multimodal} introduce KDD-Net, a model that uses two teacher models, each of which is trained on all available data that includes samples with a complete set of modalities identified as $X^{1c}$ and $X^{2c}$, as well as samples with missing modalities identified as $X^{1u}$ and $X^{2u}$, which are samples that only contain the first and second modality, respectively. The soft labels for the samples in $X^{1c}$ and $X^{2c}$ are then generated using these two single-modal teacher networks. The student model is a multi-modal deep neural network that can learn a combined representation from the several modalities available. The student model is trained with the generated soft labels by the teacher networks and one hot label, as shown in Figure \ref{fig:KDD}.

\begin{figure}[h]
		\centering
		\includegraphics[width=0.5\textwidth]{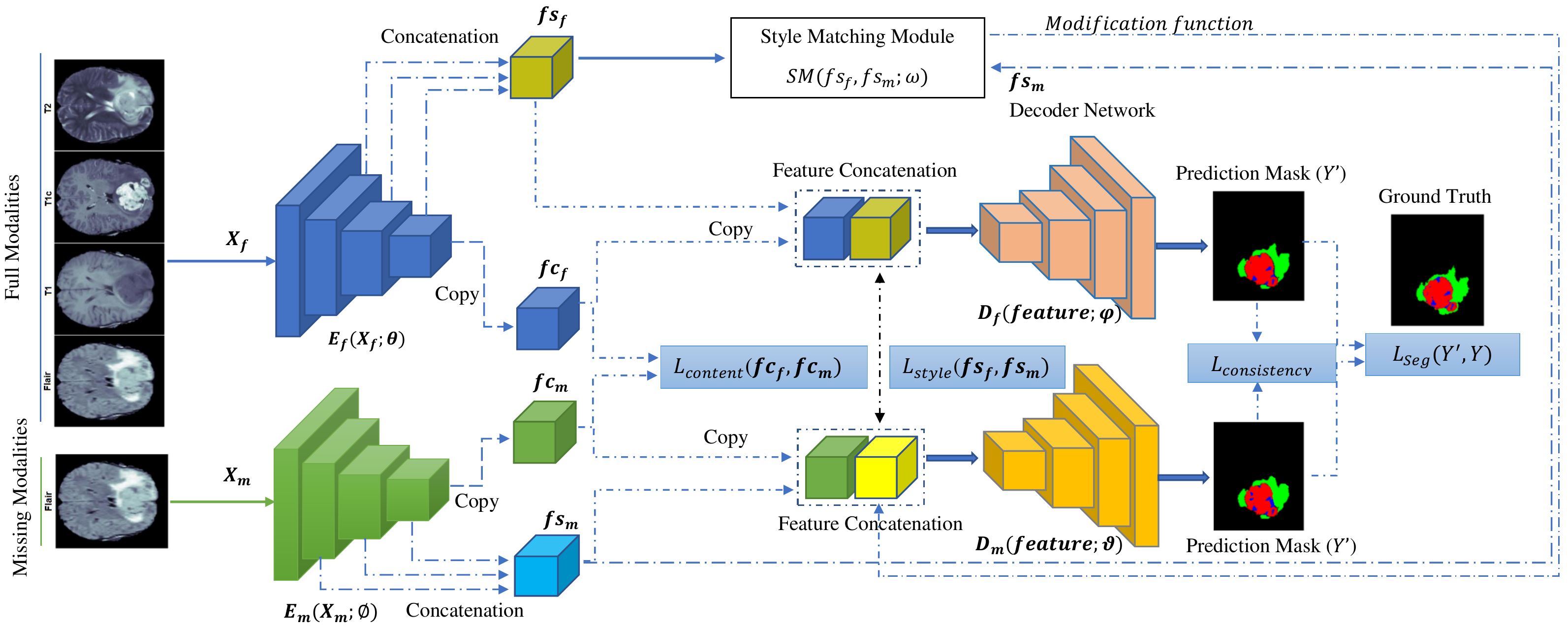}
	\caption{ The style matching method proposed in \cite{azad2021smu}. The method utilizes a knowledge distillation approach to match style and content representation between the full-modality and missing modality paths.}
	\label{fig:smunet}
\end{figure}

In another work, the style matching U-Net is proposed to overcome the problem of missing modalities. This approach builds its assumption on the idea that the feature representation in the latent space can be decomposed into a style and content representation. Then it performs both style and content matching in different levels to distill the informative features from the full-modality path into a missing modality network. The architecture of this approach is depicted in Figure\ref{fig:smunet}.

\subsection{Mutual Information Maximization}
In order to achieve minimal information loss in the missing modality situation, the mutual information maximization strategy entails optimizing similarity metrics between available modalities during training. 

\begin{figure}[h]
		\centering
		\includegraphics[width=0.5\textwidth]{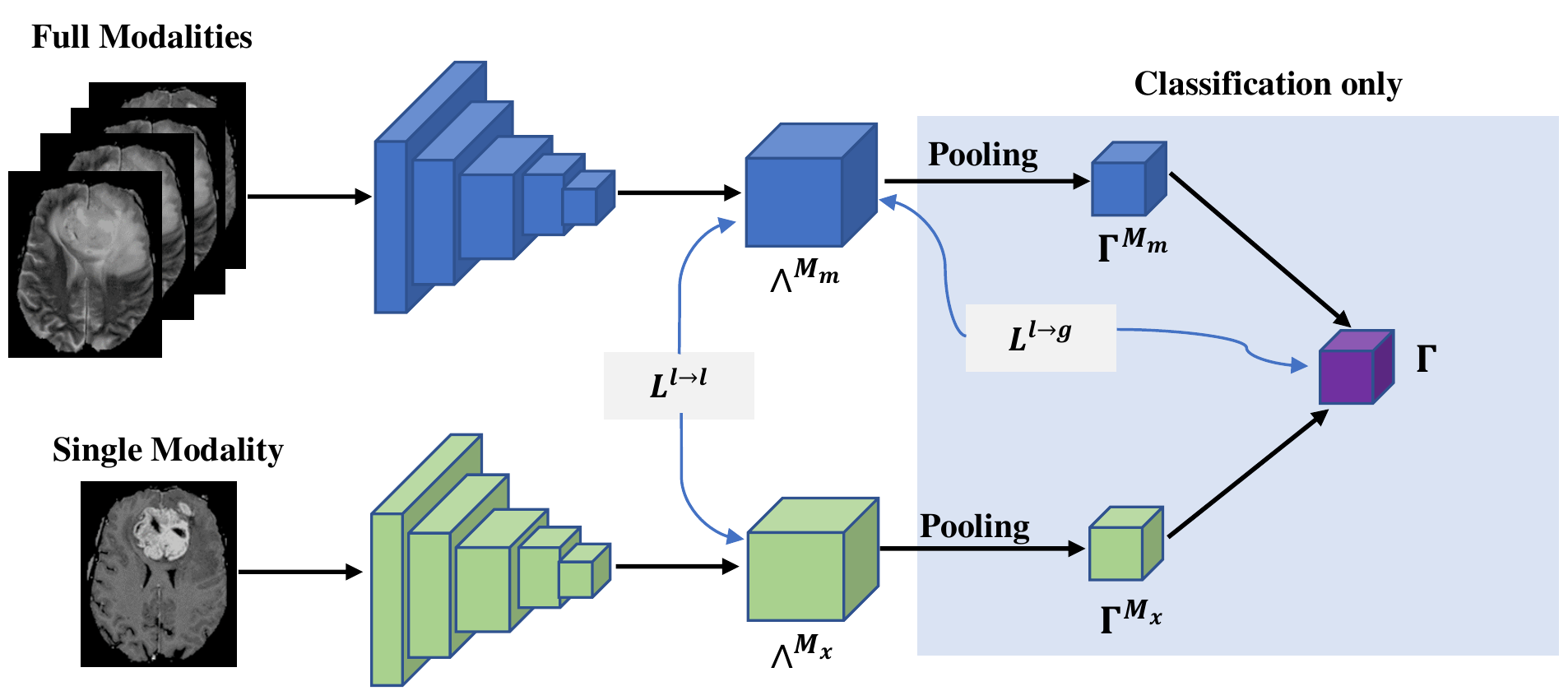}
	\caption{ CMIM architecture. The model is trained on a set of modalities $M = \{M_x,M_1,...,M_n\}$ but the only modality available at test time is $M_x$. $\Lambda^{M_i}$ is 
	 modality local feature for each modality, $\Gamma^{M_i}$ is the  modality global feature for each modality, $\Gamma$ is representing the multi-modal global embedding shared over all available modalities, $L^{l \rightarrow l}$ is the cross modal local-local loss and $L^{l \rightarrow g}$ is the cross modal local-global loss. From \cite{sylvain2020cross} }
	\label{fig:fig18}
\end{figure}

Cross-modal Information Maximization for Medical Imaging (CMIM) 
\cite{sylvain2020cross} is one the networks that utilizes the aforementioned strategy and maximizes mutual information between modalities instead of using a shared latent variable across all modalities. CMIM uses numerous  modalities for training but just one  modality at test time. The information about local and global features is then retrieved from the input images. For the semantic segmentation task, the local-local loss is computed, and for the classification task, both the local-local and local-global losses are calculated, see Figure \ref{fig:fig18}. The mutual information neural estimate approach (MINE) proposed in \cite{belghazi2018mine} is used by CMIM for mutual information estimation.

\begin{figure}[h]
\centering
\includegraphics[width=0.5\textwidth]{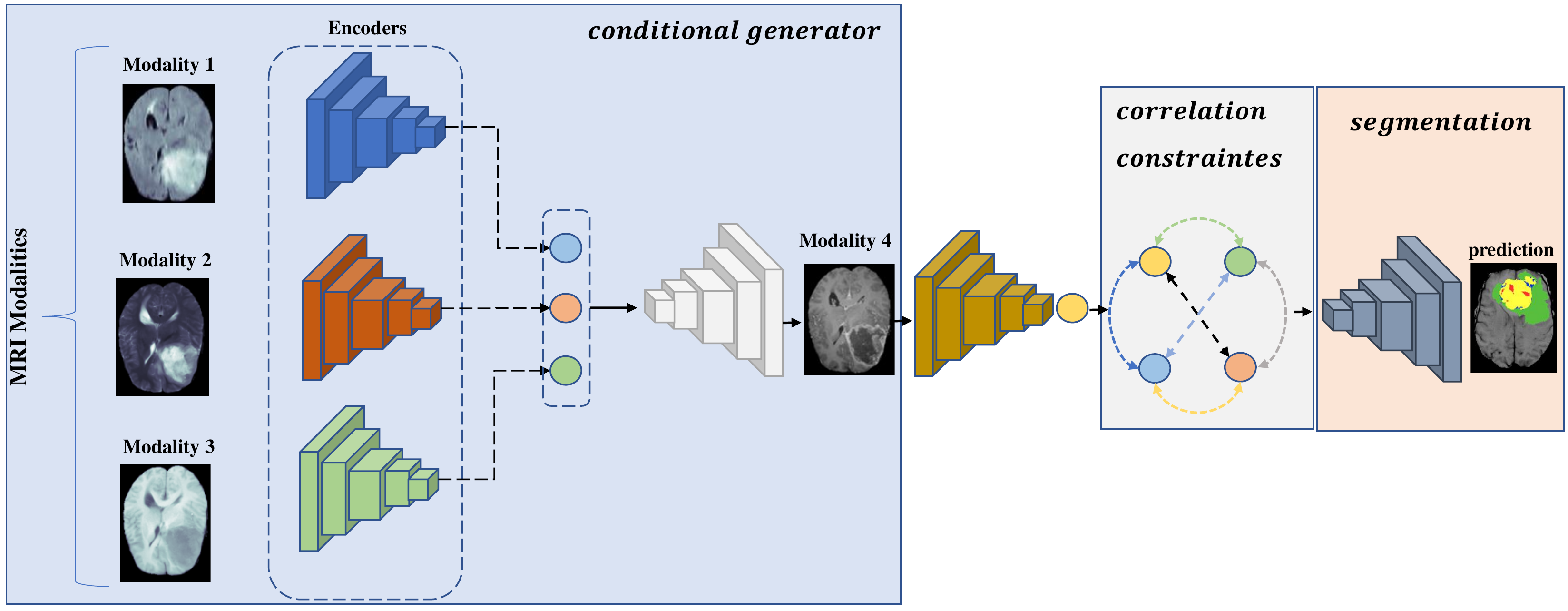}
\caption{A diagram of the suggested network from \cite{zhou2021conditional} , which includes a conditional generator, a correlation constraint network, and a segmentation network, with $X_4$ denoting the missing modality.}	\label{fig:fig19}
\end{figure}

The network described in 
\cite{zhou2021conditional} takes advantage of the significant correlation, which each two of modalities share. The architecture of this network, as also illustrated in Figure \ref{fig:fig19} is comprised of three subnetworks: The first subnetwork is a conditional generator, or more precisely, a conditional U-Net with several encoders, capable of producing the missing modality in a more monitored manner by utilizing the index of the missing modality as the condition (Each of 0, 1, 2, 3 indexes correspond to T2, T1c, FLAIR and T1 respectively). 

The authors used a correlation constraint (CC) network as their second subnetwork to compute the multi-source correlation, taking into consideration the intensity distribution profiles and their correlation. The final segmentation is determined by the third subnetwork, which is a segmentation network.

Similarly, \cite{zhou2021latent} is a latent correlation representation learning method for addressing the missing modality problem. The latent correlation representations are created when each modality is encoded individually and provided to a Model Parameter Estimation Module (MPE Module), which is then fed to a Linear Correlation Expression Module (LCE Module). The correlation model is built by the MPE and LCE modules. The latent correlation representations then travel via the fusion block, resulting in a fused representation that spans all modalities. 
\begin{figure}[h]
		\centering
		\includegraphics[width=0.5\textwidth]{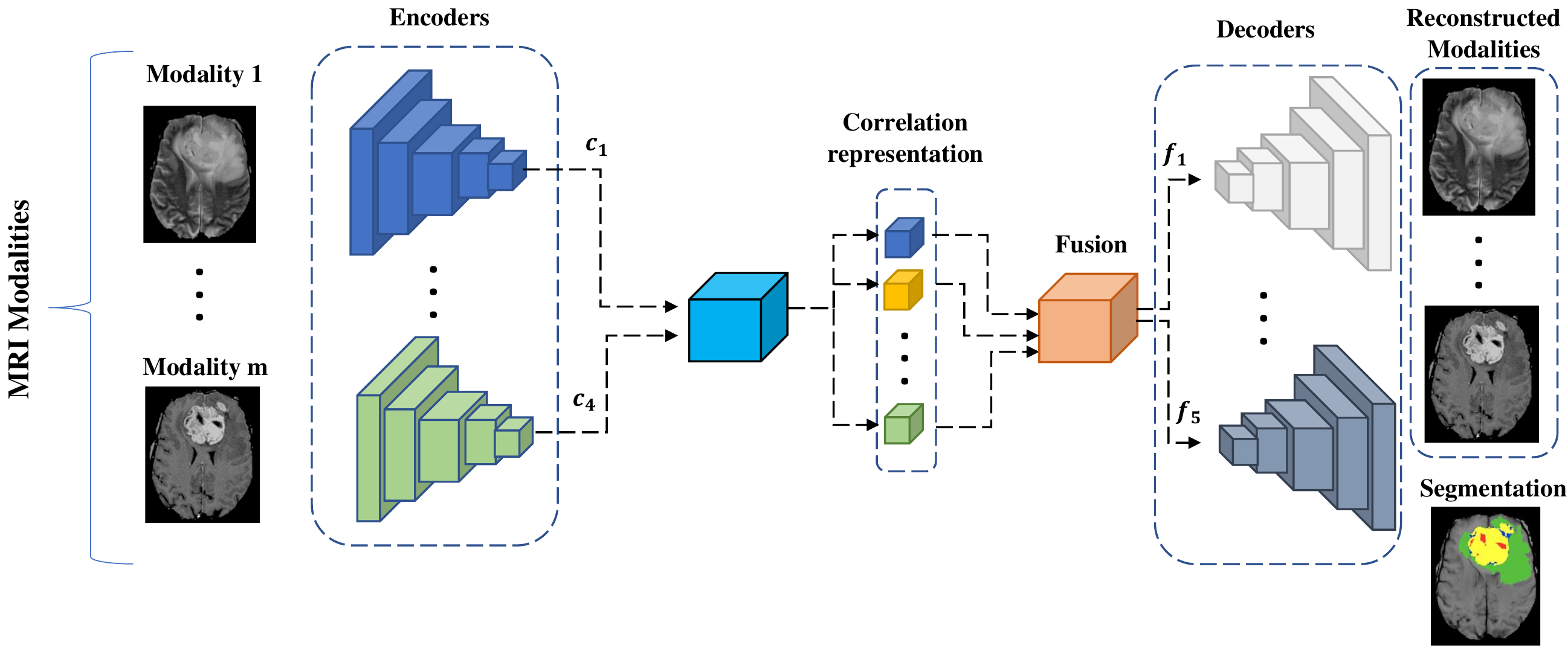}
	\caption{An illustration of the correlation and fusion method \cite{zhou2021latent}. The individual encoders are defined to produce the feature maps for each input modality. Then the correlation and fusion layers are proposed to model the underlying information and finally generate the segmentation map by reconstructing the missing information.}
	\label{fig:fig20}
\end{figure}
In the fusion block, a channel attention module and a spatial attention module are employed. By decoding the fused representation, the network will recreate the modalities and generate the segmentation, see Figure \ref{fig:fig20}.

\subsection{Generative Adversarial Networks}
GAN stands for Generative Adversarial Network and is a machine learning method initially presented in \cite{goodfellow2014generative}. 
GANs are comprised of two networks: a generator G that creates the missing modality in this case study, and a discriminator D that determines if the sample presented to it was created by the generator or is part of the original training data. Both the generator and the discriminator are trained simultaneously. This strategy will increase the generator's performance and, as a consequence, reconstruct a sample that contains and reflects the missing information. 

\begin{figure}[h]
		\centering
		\includegraphics[width=0.5\textwidth]{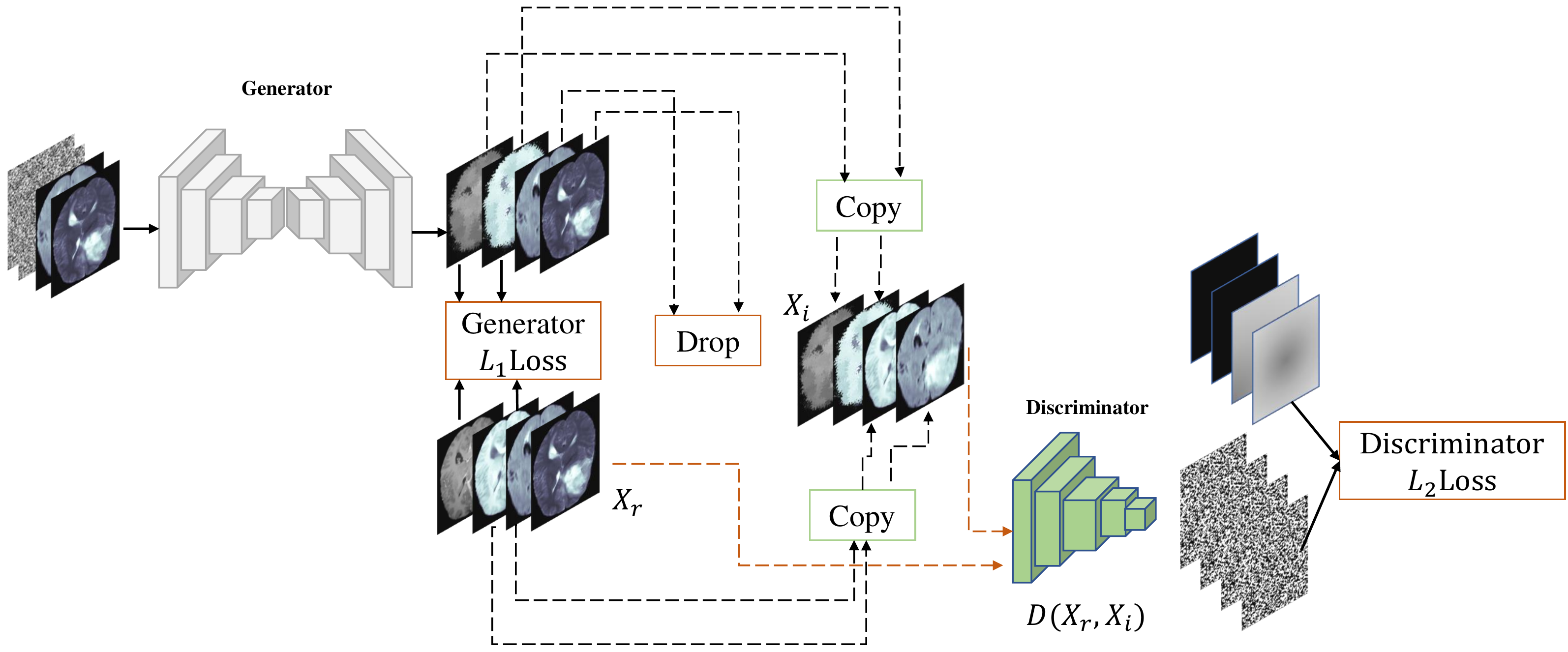}
	\caption{The Multi-Modal Generative Adversarial Network (MM-GAN) \cite{sharma2019missing}.}
	\label{fig:fig22}
\end{figure}

The authors of \cite{sharma2019missing} offered a generative adversarial network (GAN) modification that synthesizes the missing modality in a single forward pass, by just using one trained model, from the multiple inputs provided. Figure \ref{fig:fig22} demonstrates the proposed multi-modal generative adversarial network (MM-GAN) that leverages implicit conditioning (IC) to enhance synthesis outcomes as follows: 1)  A U-Net as Generator first imputes the missing modality, 2) The $L1$ loss then is determined for the scans produced by the generator, 3) The discriminator is a PatchGAN \cite{isola2017image} with modality-selective $L2$ loss computation (least squares GAN).

\begin{figure}[!h]
		\centering
		\includegraphics[scale = 0.4]{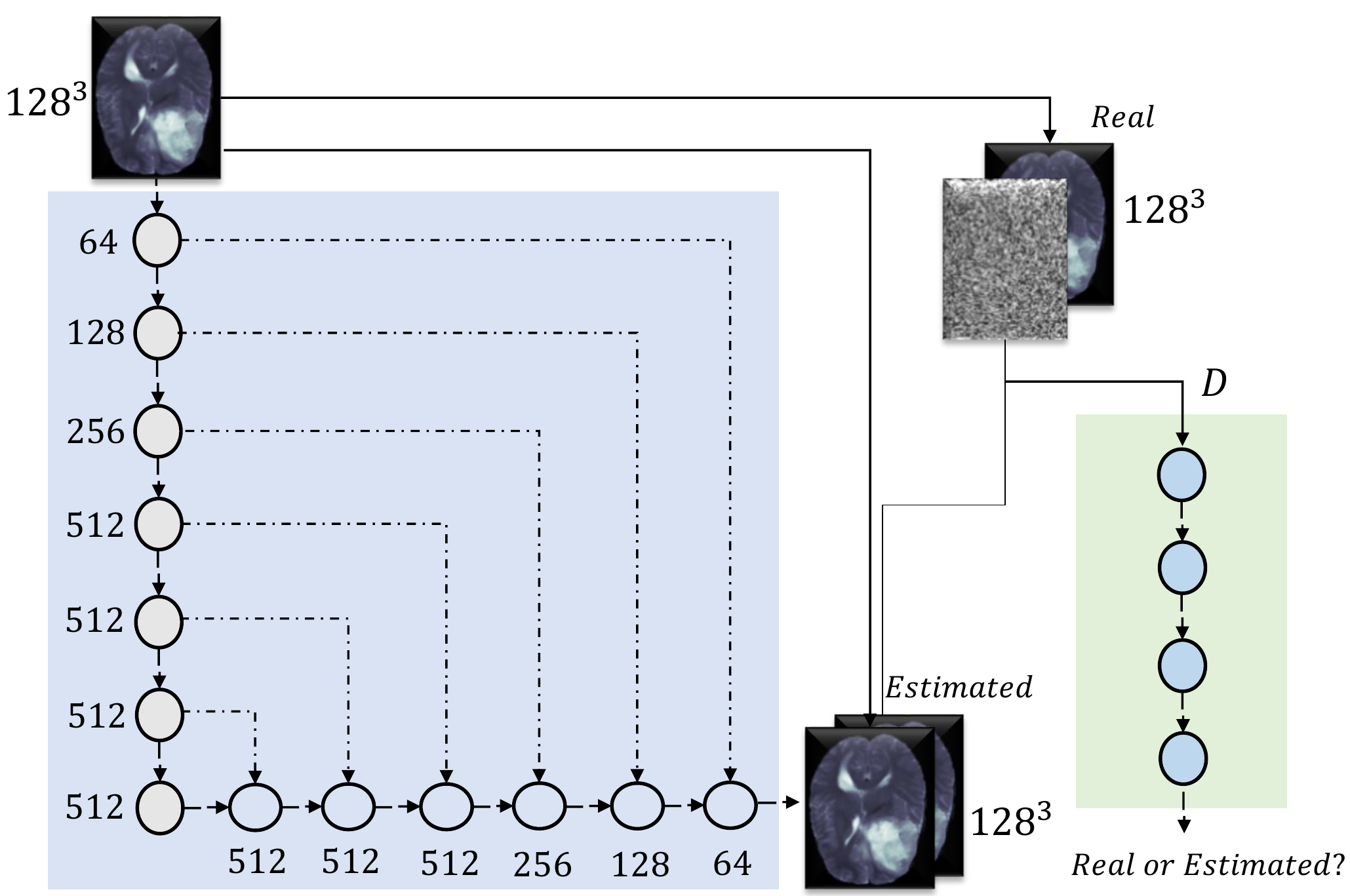}
	\caption{The 3D conditional Generative Adversarial Network \cite{yu20183d}.}
	\label{fig:fig21}
\end{figure}

Using conditional Generative Adversarial Networks (cGAN) to cope with the missing modality issue is another alternative available. CGAN is a traditional GAN extension that collects additional information In both generator and discriminator. The 3D cGAN proposed in \cite{yu20183d} synthesizes the FLAIR sequence only from T1 MR images (figure \ref{fig:fig21}). However, due to their extensive training, using GANs may not be the best approach.

\subsection{Comparative Overview}
This subsection compares and contrasts missing modality compensation approaches provided in Sections 3.1 - 3.5. Table \ref{tab:tab2} highlights their key techniques and describes the benefits and drawbacks of each model to provide the reader a clearer picture. As detailed in Table \ref{tab:tab2}, the synthesis approach performs the compensation based on the image reconstruction from an atlas sample. In practice, these approaches usually fail to reconstruct the missing information and result in no performance gain. On the other hand, the common latent space representation methods perform the information retrieval by modelling joint information from all modalities. However, these methods do not perform well when more than one modality is missing. \\
In the third direction, the knowledge distillation methods are proposed to use the strength of the co-training approach and distill the informative feature from a full-modality path into a missing modality network. Although the knowledge distillation can encourage feature learning by the student network, important domain knowledge from the full-modal network is usually not gained by the student model. Hence, mutual information matching seems to be a necessary factor to be included in the matching pipeline. \\
To this end, the fourth direction uses mutual information maximization algorithms. This approach calculates similarity metrics across available modalities and optimizes the mutual information. Regardless of the strong matching gain that can be derived from these methods, when an insufficient number of modalities are available, applying this strategy will not necessarily assure that the lost data is recovered since there won’t be enough features to reconstruct the missing data. \\
In the last strategy, the GAN methods are utilized along with the segmentation networks to compensate for the missing information. It should be noted that the extensive training costs and noisy synthesis resuls are the natural weaknesses of the GAN methods, which might result in an unstable reconstruction. \\

\begin{table*}[!ht]
\centering
\caption{A comparative overview between networks that compensate for missing modalities.}\label{tab:tab2}
\resizebox{\textwidth}{!}{
\begin{tabular}{|| p{0.1\textwidth} | p{0.4\textwidth} | p{0.2\textwidth} | p{0.3\textwidth} ||} 
 \hline
 \multicolumn{1}{||c}{\textbf{Strategy}} & \multicolumn{1}{c}{\textbf{Networks}} & \multicolumn{1}{c}{\textbf{Novelty} }& \multicolumn{1}{c||}{\textbf{Weakness}} \\ [0.5ex] 
 \hline\hline
 Synthesis Models & ``why does synthesized data improve
    multi-sequence classification'' \cite{van2015does}
    \newline
    REPLICA \cite{jog2017random}
    \newline
    mri-based attenuation correction for PET/MRI: a novel approach combining pattern recognition and atlas registration'' \cite{hofmann2008mri} & Using a more adaptable synthesis method such as NN or RBM might result in performance enhancement, \newline Uses atlas registration methods  
    & Results in no improvement in case of utilizing a classifier in a not well adjustable model framework \cite{van2015does}. \newline The majority of these models do not alter the downstream tasks such as segmentation \cite{lau2019unified}. \newline When using uniform atlases derived from healthy persons for glioma patients, distortion occurs \cite{yu20183d}. \\ 
 \hline
 Common Latent Space Models &  HeMIS \cite{havaei2016hemis} \newline
    PIMSS \cite{varsavsky2018pimms}\newline
    RS-Net \cite{mehta2018rs}\newline
    ``Brain Tumor Segmentation on
    MRI with Missing Modalities'' \cite{shen2019brain}\newline
    HVED \cite{dorent2019hetero}\newline
    ``Anatomy-Regularized Representation Learning for Cross-Modality Medical Image Segmentation''\cite{chen2020anatomy}\newline
    ACN \cite{wang2021acn}\newline
    RFNet \cite{ding2021rfnet}\newline
    URN \cite{lau2019unified}
     & Maps the available modalities into a common latent subspace and aims to recover the missing information using the newly built latent representation & Unable to adequately recover the lost information by using methods such as computing the first and second moments. \newline When more than one modality is lacking, many of these networks operate inadequately \cite{wang2021acn}. \newline They usually fail to be resilient to missing modalities while also delivering an accurate segmentation \cite{sharma2019missing}.  \\
 \hline
    Knowledge Distillation Networks & HAD-Net \cite{vadacchino2021had} \newline
    KDD-Net \cite{wang2020multimodal}\newline
    ``Knowledge distillation from multi-modal to mono-modal segmentation networks'' \cite{hu2020knowledge}\newline
    SMU-Net \cite{azad2021smu}
    & Transfers discrminitive information from one or more teacher networks to a student network for recovering the missing data & Important domain knowledge from the full-modal network is usually not gained by the student model \cite{wang2021acn}.\newline Exhibits significant training expenses when working with complex and large teacher networks. \newline Capacity mismatch between teacher and student could result in no improvement in student network's outcomes \cite{azad2021smu}. 
    \\\hline

    Mutual Information Maximization & CMIM \cite{sylvain2020cross} \newline
    ``Conditional generator and multi-sourcecorrelation guided brai tumor segmentation with missing mr modalities'' \cite{zhou2021conditional}\newline
    ``Latent correlation representation learning for brain tumor segmentation with missing mri modalities'' \cite{zhou2021latent}\newline
    ``Brain graph synthesis by dual adversarial domain alignment and target graph prediction from a source graph'' \cite{bessadok2021brain}
    & Calculates similarity metrics across available modalities and optimizes the mutual information  &  When insufficient number of modalities are available, applying this strategy will not necessarily assure that the lost data is recovered since there won't be enough features to reconstruct the missing data \cite{ding2021rfnet}.  \newline Some earlier models tended to confine the network's structure \cite{sylvain2020cross}. \\\hline
    
    Generative Adversarial Networks (GANs) & MM-GAN \cite{sharma2019missing} \newline 3D conditional Generative Adversarial Network \cite{yu20183d}\newline
    ``Multi-modal AsynDGAN: Learn From Distributed Medical Image Data without Sharing Private Information''
    \cite{chang2020multi} \newline
    Auto-GAN
    \cite{cao2020auto}\newline
    DiamondGAN
    \cite{li2019diamondgan}\newline
     CoCa-GAN \cite{huang2019coca} & Employs GAN and its modifications in the missing modality model framework & Could generate undesirable imputation noise, when imputing or synthesizing the missing modality \cite{wang2020multimodal}. \newline GANs might show a non-converging nature. \newline Extensive training costs. \newline Generator might become unstable.  \\\hline
\end{tabular}
}
\end{table*}

\begin{figure*}[h]
		\centering
		\includegraphics[width=0.9\textwidth]{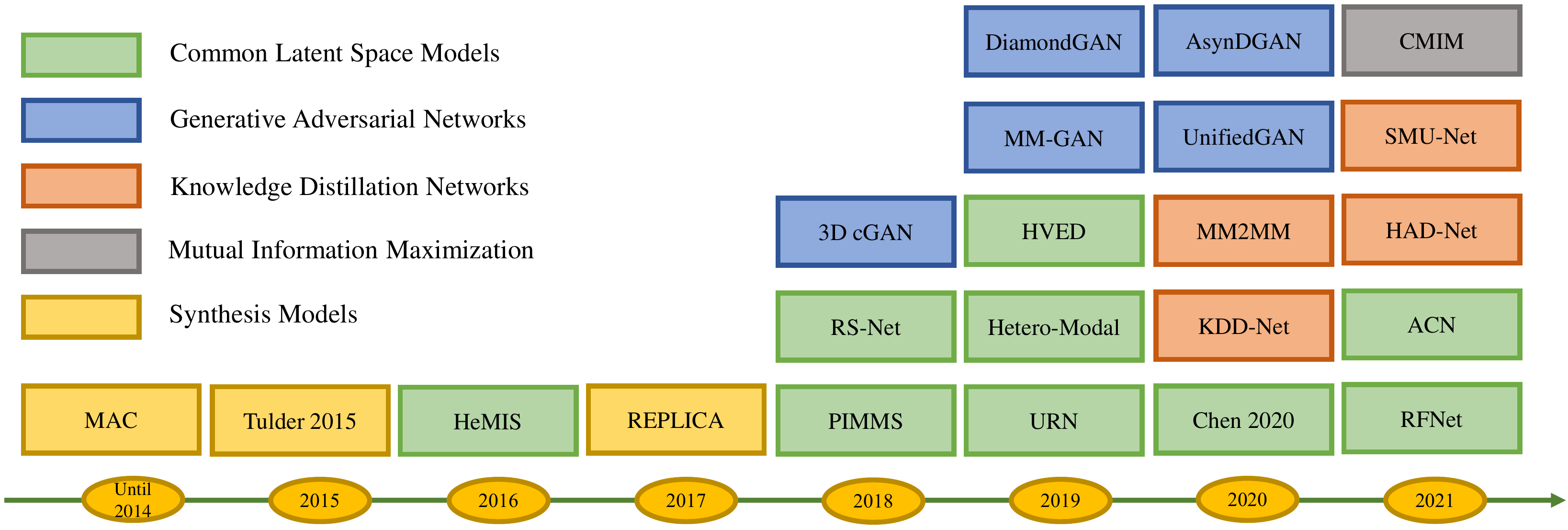}
	\caption{The timeline of the deep learning methods proposed to compensate missing modalities in MRI-based semantic segmentation, from 2014 to 2021.}
	\label{fig:figtimline}
\end{figure*}

All in all, the strategy choice for designing the network should be in accordance with the clinical application and case of study. For a more robust network, the strength of different directions should be unified into a single network. Such an approach is proposed in \cite{azad2021smu}, where the author uses the knowledge distillation approach along with the information maximization and adversarial losses. Figure \ref{fig:figtimline} demonstrates the timeline of popular deep learning approaches presented for semantic segmentation on MRI images with missing modalities since 2016. The timeline information reveals increasing attention to the missing modality challenge in recent years due to the applicability of these approaches in clinical applications. 

\section{Dataset}
In this section, we will introduce a summary of the most common MRI datasets for the task of semantic segmentation. These datasets include pixel-wise annotation to evaluate model performance. Some articles use data augmentation to increase the amount of annotated data, especially when dealing with small amounts of annotated data. Data augmentation increases the amount of training data by applying various transformations to images, which can be directly applied to images, feature space, or both. Some typical examples of these transformations include rotation, translation, scale, color jittering, cutting, and warping. In medical images, which we usually deal with a small number of images, data augmentation helps to better train models. Other benefits of data enhancement include: prevents overfitting, better generalization as well as faster convergence.

\subsection{BraTS}
\begin{figure}[h]
		\centering
		\includegraphics[width=0.5\textwidth]{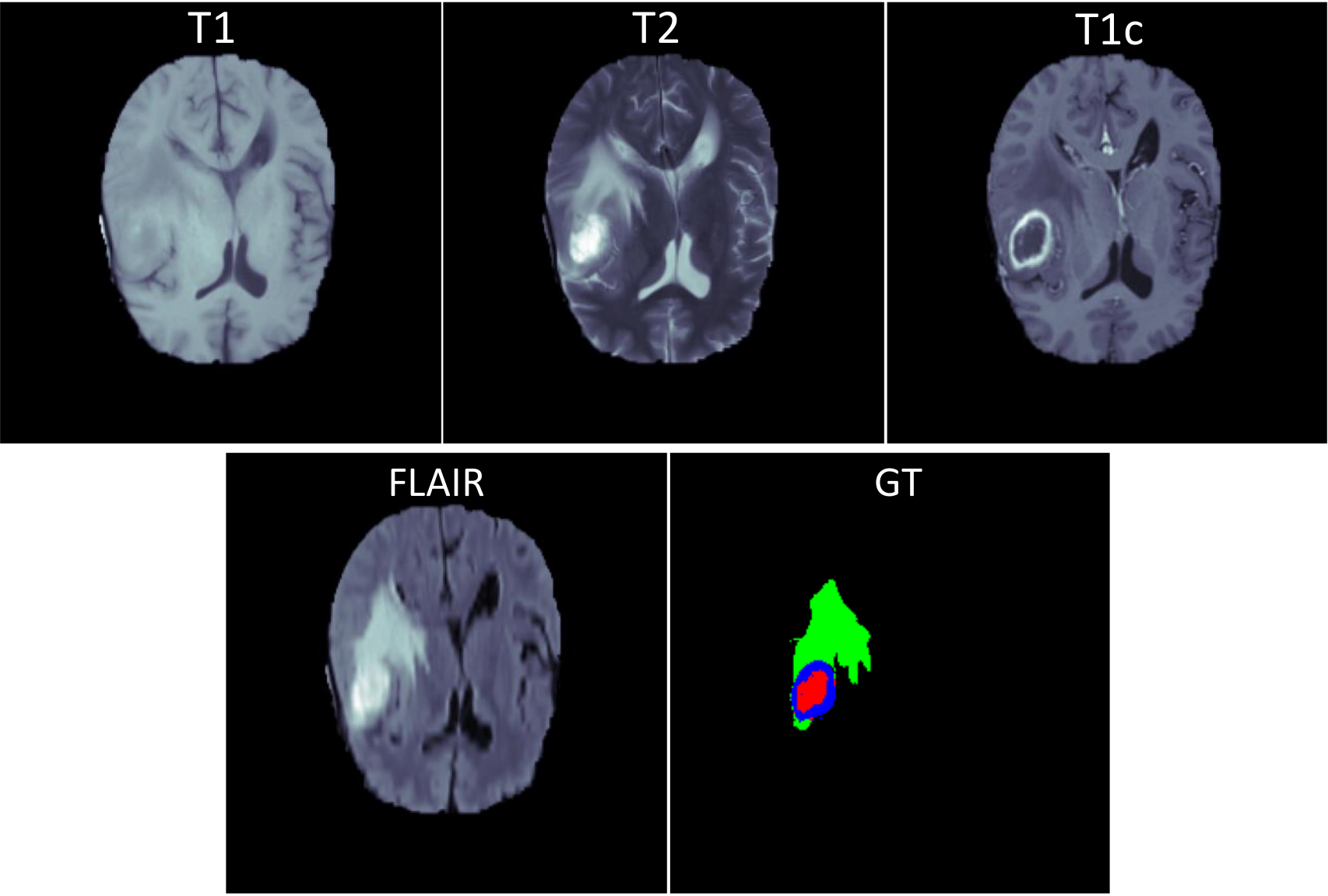}
	\caption{ Sample image from BraTS dataset \cite{myronenko20183d}. In the annotation mask the Blue area shows: GD-enhancing tumor, Green: the peritumoral edema, and Red: tumor core.}
	\label{fig:bratsdataset}
\end{figure}

The Multimodal Brain Tumor Image Segmentation Benchmark (BraTS) \cite{myronenko20183d,menze2014multimodal} is a freely accessible dataset that comprises manually segmented images provided by clinical specialists from various institutes. It was initially released in 2012 and has since then been extensively used to improve automatic segmentation methods. The BraTS dataset concentrates on gliomas, a heterogeneous group of tumors that are one of the most frequent kinds of primary brain tumors. 
T1-weighted, contrast enhanced T1 weighted also known as T1c-weighted, T2-weighted, and Fluid Attenuation Inversion Recovery (FLAIR) sequences are the four MRI sequences, which are included in the BraTS dataset, each of which reveals distinct characteristics of the brain tumor. 
For the evaluation process different version of BraTS dataset has been used by the literature work. Each verion of this dataset includes MRIs of a several patients with four modalities (T1, T1c, T2, FLAIR). Each image's ground truth segmentation in the BraTS  dataset includes labeling for four tissue classes: necrosis, edema, non-enhancing tumor, and enhancing tumor. Despite the fact that four distinct tumor labels are provided, they could well be categorized into three subregions for evaluation: the whole tumor (WT), the core tumor (CT), and the enhancing tumor (ET). Please refer to table 2 for more details on each version of BraTS dataset. Sample of image from BraTS dataset is depicted in Figure \ref{fig:bratsdataset}.

\subsection{MSGC}
The Multiple Sclerosis Grand challenge (MSGC) \cite{styner20083d} dataset consists of MR scans of 43 subjects from Boston Children’s Hospital (CHB) and the University of North Carolina (UNC) with T1, T2, FLAIR, Diffusion Tensor Imaging (DTI), and Mean Diffusivity (MD) images acquired. There are 20 training images with manual ground truth lesion segmentation and the other 23 testing images in which lesions masks are not available. T  here is an automated system to evaluate the output of the segmentation network.

\subsection{RRMS}
The Relapsing-Remitting Multiple Sclerosis (RRMS) \cite{havaei2016hemis} dataset contains MRI scans of 300 RRMS patients in the sagittal plane, including T1, T2, and T1c modalities.

\subsection{ADNI}
The Alzheimer's Disease Neuroimaging Initiative (ADNI) \cite{varsavsky2018pimms} database comprises MRI scans of 973 Alzheimer's patients, including T1 and FLAIR sequences, captured using scanners from only three manufacturers: GE, Philips, and Siemens. It's also relevant to note that ADNI suggests a unified and specific criteria for the scans it contains, resulting in a lack of diversity across its containing scans. 

\begin{table*}[!ht]
\centering
\caption{Most common datasets utilized in the literature work to evaluate the performance of missing modality compensation networks.}
\resizebox{\textwidth}{!}{
\begin{tabular}{||p{0.2\textwidth}| p{0.2\textwidth}| p{0.2\textwidth}| p{0.4\textwidth}||} 
 \hline
 \multicolumn{1}{||c}{\textbf{Datasets}} &
 \multicolumn{1}{|c}{\textbf{Num Samples}}&
 \multicolumn{1}{|c}{\textbf{Modalities}} & 
 \multicolumn{1}{|c||}{\textbf{articles}} \\ [0.5ex] 
 \hline\hline
 Multiple Sclerosis Grand challenge (MSGC) \cite{styner20083d} & 43 subjects: 20 training, 23 testing & T1, T2, FLAIR & \cite{havaei2016hemis} \\ 
 \hline
 Relapsing Remitting Multiple Sclerosis (RRMS) \cite{havaei2016hemis} & 300 subjects & T1, T2, T1C & \cite{havaei2016hemis} \\
 \hline
 BRATS2013 \cite{menze2014multimodal} &  30 subjects: 20 high grade and 10 high grade & T1, T1C, T2, FLAIR & \cite{van2015does}  \\
 \hline
 BraTS2015 \cite{menze2014multimodal} & 274 subjects: 220 high grade and 54 low grade tumors) & T1, T1C, T2, FLAIR &  \cite{mehta2018rs}; \cite{ding2021rfnet}; \cite{havaei2016hemis}; \cite{sylvain2020cross}; \cite{yu20183d}; \cite{giacomello2019transfer}, \cite{chen2019robust}; \cite{dalmaz2021resvit}\\
 \hline
  BraTS2017 \cite{menze2014multimodal}& 285 subjects: 210 high grade and 75 low grade tumors & T1, T1C, T2, FLAIR &  \cite{mehta2018rs}, \cite{shen2019brain}; \cite{islam2021glioblastoma}; \cite{dalmaz2021resvit} \\
  \hline
 BraTS2018 \cite{myronenko20183d} & 285 subjects: 210 high grade and 75 low grade tumors & T1, T1C, T2, FLAIR &  \cite{lau2019unified}; \cite{dorent2019hetero}; \cite{cao2020auto}; \cite{zhou2020hi}; \cite{chang2020multi};\cite{wang2021acn}; \cite{ding2021rfnet},  \cite{zhou2021conditional}; \cite{zhou2021latent}; \cite{sharma2019missing}; \cite{zhu2021brain}; \cite{vadacchino2021had}; \cite{zhang2021modality}\\ 
   \hline
     BraTS2019 \cite{myronenko20183d}& 335 subjects: 259 high grade and 76 low grade tumors & T1, T1C, T2, FLAIR &  \cite{vadacchino2021had}; \cite{zhou2021latent}; \cite{hamghalam2021modality}; \cite{dalmaz2021resvit}  \\
  \hline
 BraTS2020 \cite{myronenko20183d}& 369 subjects & T1, T1C, T2, FLAIR &  \cite{ding2021rfnet} \\
 \hline
 Alzheimer’s Disease Neuroimaging Initiative (ADNI) \cite{jack2008alzheimer} & 973 subjects & T1, FLAIR &  \cite{varsavsky2018pimms}; \cite{wang2020multimodal} \\
  \hline
 SABRE \cite{tillin2012southall}& 586 subjects with same scanner, 1263 subjects with multiple scanners and multiple settings & T1, T2, FLAIR & \cite{varsavsky2018pimms} \\
  \hline
MICCAI-WMH dataset \cite{web:lang:stats5} & 60 subjects & T1, FLAIR &  \cite{varsavsky2018pimms,orbes2018simultaneous,li2019diamondgan} \\
 \hline
 Ischemic Stroke Lesion Segmentation Challenge 2015 (ISLES2015) \cite{maier2017isles}& 1) SISS: 28 training and 36 testing
2) SPESS: 30 training and 20 testing
 & 1) FLAIR, T2 TSE, T1 TFE/TSE, DWI
2) T1C, T2, DWI, CBF, CBV, TTP, Tmax
 & \cite{sharma2019missing} \\
 \hline
  CHAOS2019 \cite{kavur2021chaos}(Combined Healthy Organ Segmentation) &  -
 & T1, T2 & \cite{yuan2020unified} \\
 \hline
 Brain tumor segmentation in medical segmentation decathlon \cite{simpson2019large}&  750 subjects: 484 training, 266 testing
 & T1, T1C, T2, FLAIR & \cite{yuan2020unified} \\
 \hline
MS lesions \cite{commowick2021multiple} &  65 subjects & Flair, T1, T2, double inversion recovery (DIR), T1C & \cite{li2019diamondgan} \\
 \hline
 IXI dataset \cite{brudfors2019empirical} &  600  subjects & T1 , T2, PD & \cite{brudfors2019empirical} \\
 \hline

\end{tabular}
}
\end{table*}

\subsection{SABRE}
The T1, T2, and FLAIR modalities are included in the Southall and Brent Revisited (SABRE) \cite{tillin2012southall} dataset, which represents two longitudinal cohorts, one with low variation across images obtained from 586 participants and the other with high variation across images received from 1263 patients.

\subsection{WMH}
The MICCAI 2017 White Matter Hyperintensity (WMH) \cite{web:lang:stats5} dataset includes 60 sets of brain MRIs, encompassing T1 and FLAIR modalities, with manual WMH annotations from three different institutes.

\subsection{ISLES2015}
The Ischemic Stroke Lesion Segmentation Challenge 2015 (ISLES2015) \cite{maier2017isles} dataset provides multi-spectral MRIs of stroke lesions in two different settings: Sub-Acute Stroke Lesion Segmentation (SISS) and Stroke Perfusion Estimation (SPES). The MRI sequences are stripped of their skulls, strictly co-registered with the FLAIR (SISS) and T1 (SPES) sequences, and re-sampled to a precise isotropic spacing for each setting.

\subsection{CHAOS2019}
The CHAOS benchmark \cite{kavur2021chaos}, or Combined Healthy Abdominal Organ Segmentation, is comprised of two databases, one containing CT scans and the other MRI images. The latter is the case of most interest in this study, thus we devote this section to it. The MRI database comprises 120 Digital Imaging and Communications in Medicine (DICOM) datasets, including 40 T1-DUAL in phase datasets, 40 T2-SPIR datasets, and 40 T1-DUAL out phase datasets, all of which were obtained utilizing different RF Pulse and gradient combinations. All the MRI scans were acquired with a 1.5T Philips machine.

\subsection{MS Lesion}
The Ms Lesion dataset \cite{commowick2021multiple} includes Only 65 scans of pathological brain MRIs of individuals with Multiple Sclerosis (MS) lesions, which contains T1, T2, FLAIR, Double Inversion Recovery (DIR), and T1c modalities.

\subsection{IXI dataset }
This dataset \cite{brudfors2019empirical} comprises almost 600 MR scans from healthy subjects, including T1, T2, and PD sequences as well as Diffusion-weighted (DW) sequences, taken on two different vendor systems: a Philips 1.5 and 3T system as well as a GE 1.5T system.
\\
Table 2 summarizes the aforementioned dataset along with the number of samples, modalities and research work performed on these benchmarks.

\section{Performance Review}

In this section, first, a summary of popular metrics in the evaluation of medical image segmentation networks will be presented, and then the quantitative performance of recent methods in the segmentation of medical images along with missing modalities will be discussed.

\subsection{Metrics For Evaluating the Performance}

Most articles in recent years have focused only on the issue of quantitative accuracy of the model and compare and report the performance of their model in terms of quantitative accuracy. They lack to includes other important aspects such as speed (inference time) and the amount of memory required (which we will discuss in section 6). In this section, we briefly introduce some popular metrics used for evaluating the accuracy of missing modality compensating networks. The results will compare the most promising method for the popular datasets.
\\
\textbf{Dice score} \\
In semantic segmentation, the Dice loss which is based on dice coefficient similarity is well-known. In the segmentation of medical images, most of the time the region of interest (ROI) is a small part of the image. Therefore, the model is prone to be trapped in the local minimum in the training process of the model. Accordingly, the model will bias to the background, the object of interest will not be detected appropriately and so many of them will miss. Thus, the Dice loss was proposed to alleviate this problem \cite{milletari2016v,azad2021texture}. The Dice loss is formulated for a 3D MRI image as written in Equation \ref{eqn:Dice_loss}:

\begin{equation}
D= 1- \frac{2 \sum_{i}^{N} p_{i} g_{i}}{\sum_{i}^{N} p_{i}^{2}+\sum_{i}^{N} g_{i}^{2}}
\label{eqn:Dice_loss}
\end{equation}

Where N is the number of voxels, $p_{i}$ is the predicted binary segmentation volume, and $g_{i}$ is the ground truth binary volume. 
\\
\textbf{Hausdorff Distance}\\
Hausdorff distance is a common performance evaluation criterion in assessing the distance between two sets of points. In fact, it is the longest distance from one point in one set to the nearest point in another set. This criterion has an advantage over other performance evaluation criteria such as the Dice score due to the consideration of voxel location. The Hausdorff distance between two sets of points A and B is calculated as written in Equation \ref{eqn:H_loss}:
\begin{equation}
h(A, B)=\max _{a \in A}\left\{\min _{b \in B}\{d(a, b)\}\right\}
\label{eqn:H_loss}
\end{equation}

\textbf{Volume Difference (VD):}
The volume difference represents the absolute percentage of the volume difference between the prediction and the ground truth. The volume difference is calculated as written in equation \ref{eqn:vd_loss}:
\begin{equation}
VD = \frac{GT - P}{GT}
\label{eqn:vd_loss}
\end{equation}
which GT and P are the ground truth and the predicted volume respectively.

\textbf{Surface Distance (SD):}
The surface distance determines the difference between the surface of the segmented object and the ground truth in three dimensions. After determining the boundary voxels of the segmentation and the ground truth, those voxels that have at least one neighbor from a predefined neighborhood that does not belong to the object are collected. For each collected voxel, the nearest voxel in the other set is determined. The surface distance is calculated as written in Equation \ref{eqn:sd_loss}:
\begin{equation}
SD = \sum_{\mathrm{S}_{\mathrm{gt}} \in \mathrm{S}(\mathrm{GT})} \mathrm{d}\left(\mathrm{S}_{\mathrm{gt}}, \quad \mathrm{S}(\mathrm{P})\right)+\sum_{\mathrm{S}_{\mathrm{p}} \in S(\mathrm{P})} \mathrm{d}\left(\mathrm{S}_{\mathrm{p}}, \quad \mathrm{S}(\mathrm{GT})\right)
\label{eqn:sd_loss}
\end{equation}

\textbf{Precision and Recall}
Prevision demonstrates the number of correct instances ($TP$) relevant from the retrieved true instances \cite{asadi2020multi}.  While the recall metric measures the fraction of $TP$ that was retrieved \ref{eqn:precall}:
\begin{equation}
Precision = \frac{TP}{TP+FP}\\
Recall = \frac{TP}{TP+FN}
\label{eqn:precall}
\end{equation}

\subsection{Quantitative Performance Analysis}
In this section, the performance of the reviewed methods on the most common benchmarks including (MSGC, MICCAI-WMH, BraTS2015 and BraTS2018 dataset) will be reported. In our comparison tables, we only include methods that are using the same setting on a particular dataset (to provide a fair evaluation). We further provide experimental results on a BraTS series with extreme missing modalities to provide a user a clear view of the overall performance gained till now. Table \ref{table3} demonstrates the experimental results on the MSGC dataset.

\begin{table}[!h]
\caption{Experimental results on the MSGC dataset. All method are compared in term of true positive (TP), false positive (FP), volume difference (VD) and surface distance (SD) metrics for both CHB and UNC raters.}\label{table3}
\centering
\resizebox{0.5\textwidth}{!}{
\begin{tabular}{l||cccccc}
\hline Method & Rater & VD (\%) & SD (mm) & TPR (\%) & FPR (\%) & Score \\
\hline \multirow{2}{*}{ \cite{souplet2008automatic}} & CHB & $86.4$ & $8.4$ & $58.2$ & $70.6$ & \multirow{2}{*}{ $80.0$} \\ & UNC & $57.9$ & $7.5$ & $49.1$ & $76.3$ & \\

\hline \multirow{2}{*}{ \cite{geremia2013spatially}} & CHB & $\mathbf{5 2 . 4}$ & $\mathbf{5 . 4}$ & $59.0$ & $71.5$ & \multirow{2}{*}{ 82.1 } \\ & UNC & $\mathbf{4 5 . 0}$ & $\mathbf{5 . 7}$ & $51.2$ & $76.7$ & \\

\hline \multirow{2}{*}{ \cite{brosch2015deep}} & CHB & $63.5$ & $7.4$ & $47.1$ & $\mathbf{5 2 . 7}$ & \multirow{2}{*}{$\mathbf{84.0}$} \\
& UNC & $52.0$ & $6.4$ & $\mathbf{5 6 . 0}$ & $\mathbf{4 9 . 8}$ & \\
 \hline \multirow{2}{*}{\cite{havaei2016hemis}}& CHB & $127.4$ & $7.5$ & $\mathbf{66.1}$ & $55.3$ & \multirow{2}{*}{$83.2$} \\
& UNC & $68.2$ & $6.6$ & $52.3$ & $61.3$ & \\
\hline
\end{tabular}
}
\end{table}

The experimental results reported in \cite{havaei2016hemis} suggest that the HeMIS method outperforms other competitors when subjects with missing modalities are presented. Table 4 focuses on the comparison results reported on the MICCAI-WMH dataset and provides a experimental results have been done by PIMMS \cite{varsavsky2018pimms} method to overcome the issue of missing labels in MRI series.

\begin{table}[h!] 
	\caption{Performance evaluation on the MICCAI-WMH dataset with missing labels.}\label{tab4}
\centering
	\resizebox{0.5\textwidth}{!}{
		\begin{tabular}{c||c||c}
			\hline
			{\begin{tabular}{ccc}
					\multicolumn{3}{c}{\textbf{Modalities}} \\
					\hline
					\textbf{T1} &  \textbf{T2} & \textbf{F} \\
					\hline
					$\bullet$ & $\bullet$ & $\bullet$ \\
					$\bullet$ & $\bullet$ & $\circ$   \\
					$\circ$   & $\bullet$ & $\bullet$ \\
					$\bullet$ & $\circ$   & $\bullet$ \\
					$\bullet$ & $\circ$   & $\circ$ \\
					$\circ$   & $\bullet$ & $\circ$ \\
					$\circ$   & $\circ$   & $\bullet$\\

				\end{tabular}
			} &
			{\begin{tabular}{cccc}
					\multicolumn{4}{c}{\textbf{Dice Score}} \\
					\hline
					\textbf{HeMIS} & \textbf{Soft} & \textbf{Hard}&\textbf{Online} \\
					\hline
                    $0.47$ & $\mathbf{0 . 5 1}$ & $0.48$ & $\mathbf{0 . 5 4}$ \\
                    $0.3$ & $\mathbf{0 . 3 9}$ & $0.3$ & $0.24$ \\
                    $0.26$ & $\mathbf{0 . 3 2}$ & $\mathbf{0 . 2 6}$ & $\mathbf{0 . 4}$\\
                    $0.44$ & $\mathbf{0 . 4 5}$ & $\mathbf{0 . 4 5}$ & $\mathbf{0 . 5 2}$\\
                    $0.1$ & $\mathbf{0 . 1}$ & $0.1$ & $\mathbf{0 . 1 9}$\\
                    $0.08$ & $\mathbf{0 . 0 8}$ & $0.07$ & $\mathbf{0 . 0 9}$ \\
                    $0.16$ & $\mathbf{0 . 1 8}$ & $\mathbf{0 . 1 6}$ & $\mathbf{0 . 4 1}$ 
					
				\end{tabular}
			} &
			{\begin{tabular}{cccc}
					\multicolumn{4}{c}{\textbf{Avg. Symmetric Distance}} \\
					\hline
					\textbf{HeMIS} & \textbf{Soft} & \textbf{Hard} & \textbf{Online} \\
					\hline
                    $0.71$ & $0.65$ & $0.71$ & $1.9$ \\
                    $2.32$ & $\mathbf{1 . 9 2}$ & $2.36$ & $4.21$ \\
                    $0.77$ & $0.82$ & $0.76$ & $3.32$ \\
                    $0.61$ & $0.63$ & $0.62$ & $2.06$ \\
                    $3.42$ & $3.76$ & $3.51$ & $4.48$ \\
                    $4.07$ & $4.13$ & $4.53$ & $7.48$ \\
                    $0.56$ & $0.61$ & $\mathbf{0 . 5 4}$ & $3.31$\\ 
	
				\end{tabular}
			} \\
		\end{tabular}
	}
\end{table}

Table \ref{table4} focuses on the BraTS2015 which is more popular benchmark for missing modality compensation networks. There have been a large number of work reported their performance on this dataset, however, in this paper we only tabulated the methods which performed the evaluation through the online system provided by the BraTS challenge.  

\begin{table}[!ht] 
	\caption{Comparison results reported on the BraTS 2015 test set.}\label{table4}
\centering
	\resizebox{0.5\textwidth}{!}{
		\begin{tabular}{c||c||c}
			\hline
			{\begin{tabular}{l}
					\multicolumn{1}{c}{\textbf{Articles}} \\
					\hline
					\\
				\cite{kamnitsas2017efficient}\\	
                \cite{zhao2018deep}\\
	            \cite{zhou2018one}\\
	            \cite{chen2019robust}
				\end{tabular}
			} &
			{\begin{tabular}{ccc}
					\multicolumn{3}{c}{\textbf{Dice(\%)}} \\
					\hline
					\textbf{Complete} & \textbf{Core} & \textbf{Enhancing} \\
					\hline
                    84 & 67 & 63  \\
                    82 & $\mathbf{7 2}$ & 62\\
                    $\mathbf{8 6}$ & 71 & $\mathbf{6 4}$\\
                    84 & $\mathbf{7 2}$ & $\mathbf{6 4}$\\
					
				\end{tabular}
			} &
			{\begin{tabular}{ccc}
					\multicolumn{3}{c}{\textbf{Precision(\%) / Sensitivity(\%)}} \\
					\hline
					\textbf{Complete} & \textbf{Core} & \textbf{Enhancing} \\
					\hline
                    82/$\mathbf{89}$ & $\mathbf{8 5}/62$ & $\mathbf{6 4}/66$\\
                    84/83 & 78/$\mathbf{73}$ & 60/69\\
	                $\mathbf{8 6}/88$ & 83/68 & 61/$\mathbf{72}$\\
	                84/$\mathbf{89}$ & 80/69 & $\mathbf{6 4}/68$
				\end{tabular}
			} \\
		\end{tabular}
	}
\end{table}

Table \ref{tab6} provided experimental results on BraTS2018 with the same setting and compares four well-known approaches for missing modality compensation. The recent approach SMU-Net and ACN outperforms the baseline methods HeMIS and HVED with large margins. It is crystal clear that, with the advance of new approaches there have been a $15\%$ performance gain achieved by the recent works since the introduction of HeMIS method.

\begin{table*}[t] 
	\caption{Performance comparison of the proposed SMU-Net on the BraTS 2018 dataset using Dice metric. Note our method uses adversarial style matching module.}\label{tab6}

	\resizebox{\textwidth}{!}{
		\begin{tabular}{c||c||c||c}
			\hline
			{\begin{tabular}{cccc}
					\multicolumn{4}{c}{\textbf{Modalities}} \\
					\hline
					\textbf{Flair} &  \textbf{T1} &  \textbf{T1c} & \textbf{T2} \\
					\hline
					$\circ$ & $\circ$ & $\circ$ & $\bullet$ \\
					$\circ$ & $\circ$ & $\bullet$ & $\circ$ \\
					$\circ$ & $\bullet$ & $\circ$ & $\circ$ \\
					$\bullet$ & $\circ$ & $\circ$ & $\circ$ \\
					$\circ$ & $\circ$ & $\bullet$ & $\bullet$ \\
					$\circ$ & $\bullet$ & $\bullet$ & $\circ$ \\
					$\bullet$ & $\bullet$ & $\circ$ & $\circ$ \\
				    $\circ$ & $\bullet$ & $\circ$ & $\bullet$ \\
				    $\bullet$ & $\circ$ & $\circ$ & $\bullet$ \\
				    $\bullet$ & $\circ$ & $\bullet$ & $\circ$ \\
				    $\bullet$ & $\bullet$ & $\bullet$ & $\circ$ \\
				    $\bullet$ & $\bullet$ & $\circ$ & $\bullet$ \\
				    $\bullet$ & $\circ$ & $\bullet$ & $\bullet$ \\
				    $\circ$ & $\bullet$ & $\bullet$ & $\bullet$ \\
				    $\bullet$ & $\bullet$ & $\bullet$ & $\bullet$ \\
				    \hline
				    \multicolumn{4}{c}{\textbf{Mean}} \\
					\hline

				\end{tabular}
			} &
			{\begin{tabular}{cccc}
					\multicolumn{4}{c}{\textbf{Complete}} \\
					\hline
					\textbf{U-HeMIS} & \textbf{HVED} & \textbf{ACN}&\textbf{SMU-Net} \\
					\hline
					79.2 & 80.9  &  85.4 & \textbf{85.7} \\
					58.5 & 62.4  &  79.8 & \textbf{80.3}\\
					54.3 & 52.4  & \textbf{78.7} & 78.6\\
					79.9 & 82.1  &  87.3 & \textbf{87.5}\\
					81.0 & 82.7  &  84.9 & \textbf{86.1}\\
					63.8 & 66.8  & 79.6 & \textbf{80.3}\\
					83.9 & 84.3  & 86.0 & \textbf{87.3}\\
					80.8 & 82.2  & 84.4 & \textbf{85.6}\\
					86.0 & 87.5  &  86.9 & \textbf{87.9}\\
					83.3 & 85.5  & 87.8 & \textbf{88.4}\\
					85.1 & 86.2  & \textbf{88.4} & 88.2\\
					87.0 & 88.0  &  87.4 & \textbf{88.3}\\
					87.0 & \textbf{88.6} & 87.2 & 88.2\\
					82.1 & 83.3  &  \textbf{86.6} & 86.5\\
					87.6 & 88.8  & \textbf{89.1} & 88.9\\
					\hline
					78.6 & 80.1  & 85.3 & \textbf{85.9}\\
					\hline
					
				\end{tabular}
			} &
			{\begin{tabular}{cccc}
					\multicolumn{4}{c}{\textbf{Core}} \\
					\hline
					\textbf{U-HeMIS} & \textbf{HVED} & \textbf{ACN} & \textbf{SMU-Net} \\
					\hline
					50.5 & 54.1  & 66.8 & \textbf{67.2}\\
					58.5 & 66.7  & 83.3 & \textbf{84.1} \\
					37.9 & 37.2  & \textbf{70.9} & 69.5 \\
					49.8 & 50.4  & 66.4 & \textbf{71.8}\\
					69.1 & 73.7  & 83.2 & \textbf{85.0}\\
					64.0 & 69.7  &  83.9 & \textbf{84.4}\\
					56.7 & 55.3  & 70.4 & \textbf{71.2}\\
				    53.4 & 57.2  &  72.8 & \textbf{73.5}\\
				    58.7 & 59.7  & 70.7 & \textbf{71.2}\\
				    67.6 & 72.9  & 82.9 & \textbf{84.1}\\
				    70.7 & 74.2  & 83.3 & \textbf{84.2}\\
				    61.0 & 61.5  & 67.7 & \textbf{67.9} \\
				    72.2 & 75.6  & \textbf{82.9} & 82.5 \\
				    70.7 & 75.3  & 83.2 & \textbf{84.4} \\
				    73.4 & 76.4  & 84.8 & \textbf{87.3}\\
				    \hline
					59.7 & 64.0  &  76.8 & \textbf{77.9} \\
					\hline

				\end{tabular}
			} &
			{\begin{tabular}{cccc}
					\multicolumn{4}{c}{\textbf{Enhancing}} \\
					\hline
					\textbf{U-HeMIS} & \textbf{HVED} & \textbf{ACN} & \textbf{SMU-Net} \\
					\hline
					23.3 & 30.8  & 41.7 & \textbf{43.1}\\
					60.8 & 65.5  & 78.0 & \textbf{78.3}\\
					12.4 & 13.7  & 41.8 & \textbf{42.8}\\
					24.9 & 24.8  & 42.2 & \textbf{46.1}\\
					68.6 & 70.2  &74.9 & \textbf{75.7}\\
					65.3 & 67.0  & \textbf{75.3} & 75.1\\
					29.0 & 24.2   &42.5 & \textbf{44.0}\\
					28.3 & 30.7   &46.5 & \textbf{47.7}\\
					28.0 & 34.6   &44.3 & \textbf{46.0}\\
					68.0 & 70.3   &\textbf{77.5} & 77.3\\
					69.9 & 71.1     &75.1 & \textbf{76.2}\\
					33.4 & 34.1  &42.8 & \textbf{43.1}\\
					69.7 & 71.2   &73.8 & \textbf{75.4}\\
					69.7 & 71.1   &75.9 & \textbf{76.2}\\
					70.8 & 71.7  &78.2 & \textbf{79.3}\\
				    \hline
					48.1 & 50.1 & 60.70 & \textbf{61.8} \\
					\hline
					
				\end{tabular}
			} \\
		\end{tabular}
	}
\end{table*}

\subsection{Extreme missing modality}
In this section, we analyze the performance of several algorithms in case of extreme missing modality. More precisely we assume that on the training time all modalities are presented, however, the inference only applies to a single modality data. This extreme missing scenario provided a good benchmark to evaluate the effectiveness of different approaches to tackle the problem of missing information. Table \ref{tab7} provides experimental results on the BraTS2015 and BraTS2018 series. It is worthwhile to mention that for each modality we reported the average dice score (average of the whole, enhance and core tumour dice scores). The experimental results show that compared to the full-modality scenario, the performance dramatically decreases in a single modality case, however, recent approaches (e.g. \cite{wang2021acn} takes the strength of knowledge distillation and information maximization approaches to train a robust model to missing modality. 

\begin{table}[!h] 
	\caption{Experimental results of the literature work on BraTS series with extreme missing modality scenario.}\label{tab7}
	\centering
	
	\resizebox{.5\textwidth}{!}{
	\scriptsize
		\begin{tabular}{l||c}
			\hline
			{\begin{tabular}{c}
					\multirow{2}{*}{\textbf{Article}}\\ \\
					\hline
					 \cite{havaei2016hemis}\\ 
					 \cite{giacomello2019transfer}\\
					 \cite{chen2019robust} \\
					 \cite{sylvain2020cross}\\
					 \\
					 \cite{dorent2019hetero}\\
					 \cite{wang2021acn}\\
					 \cite{ding2021rfnet}\\
					 \cite{zhou2021conditional}\\
					 \cite{azad2021smu} \\
					 \cite{zhou2021latent}\\
					
			\end{tabular}
			} 
&
			{\begin{tabular}{cccc|c}
					\multicolumn{5}{c}{\textbf{BraTS2015}} \\
					\hline  
					\textbf{T1} & \textbf{T1c} & \textbf{T2} & \textbf{FLAIR} & \textbf{AVG}\\
					\hline
					4.67  & 49.93 & 20.31 &  5.57 & 20.12  \\
					$\mathbf{56.19}$ & 63.40 & $\mathbf{75.65}$ &  $\mathbf{80.25}$ & $\mathbf{68.87}$\\
					47.02 & $\mathbf{71.65}$ & 60.60 &  52.66 & 57.98\\
					14.15 & 49.00 & 29.56 &  23.37 & 29.02\\
				\hline
				\multicolumn{5}{c}{\textbf{BraTS2018}} \\
				\hline
					34.43 & 64.86 & 55.26 & 52.43 & 51.74\\
					81.0  & $\mathbf{82.7}$  & $\mathbf{86.1}$  &  $\mathbf{86.41}$ & $\mathbf{84.05}$\\
					60.16 & 77.71 & 67.7  & 64.88 & 67.61\\
				    $\mathbf{83.6}$  &  63.2 &  84.0 & 81.7  & 78.12\\
				    63.6  & 80.9  & 65.3  & 68.4  & 69.5 \\
				    5.7   & 47.73 & 18.7  & 49.36 & 30.37 
			\end{tabular}
			}  
		\end{tabular}}
		
\end{table}

\section{Challenges and Opportunities}
In recent years, promising deep-learning-based methods have been introduced to equip medical imaging with Missing modalities. Here some perspectives on future research will be introduced that can further improve the methods of medical image segmentation with Missing modalities.

\subsection{More Challenging Datasets}
Section 4 introduced most popular MRI datasets for semantic segmentation. The BraTS dataset, for instance, contains a large number of 3D MRI sequences but appears to lack image format variance. This lack of variety may lead us to the conclusion that more challenging datasets, and more representative of real-world situations, are needed to enhance the training process and urge the models to provide better results.

\subsection{Memory Efficient Models}
It's noteworthy to mention that the majority of the approaches discussed in this paper are primarily concerned with offsetting the negative consequences of operating with an incomplete set of MRI modalities and, as a result, increasing segmentation accuracy. However, these approaches often need a large amount of memory, not only during the training phase but also throughout the inference. Knowledge Distillation Networks, as discussed in section 3, are one of the key techniques for the issue at hand, and they may be used to transfer knowledge from a larger, more complicated model to a smaller, less memory-intensive one. A simplified network could be achieved using the knowledge distillation method or network compression techniques, which can then be deployed in other devices such as smartphones.

\subsection{Balance Between Accuracy and Efficiency}
In machine learning and deep learning models, there is a well-known trade-off between accuracy and efficiency, and semantic segmentation networks are no exception. It is frequently the case that models that are capable of producing more accurate outcomes have a lower efficiency level. This is also truly the case in the inverse scenario, implying that the efficient models are less accurate. Future work should consider this fact in their design. 

\subsection{Model complexity}
As discussed in the previous section, a small number of articles report information such as computational complexity, run-time, and memory footprint, which is important for clinics that may have limited computing resources. Besides the memory shortage in some devices, inference time plays a critical role in some real-time applications. Thus, model complexity needs to be taken into account for such a scenario. Model parameters, Floating-Point Operations (FLOPs), Runtime, and Frame Per Second (FPS) are all commonly used metrics to assess the model's complexity. Model parameters and FLOPs, the first two metrics are given, are notably independent of the implementation environment, and the larger their value, the lower the implementation efficiency. Because of their reliance on the hardware and implementation environment, runtime and FPS are two metrics for assessing implementation speed that may be deemed less favourable than model parameters and FLOPs. For a real-world application, these measurement needs to be considered. 

\subsection{Interpretable Models}
From a clinical perspective, it is highly desirable to understand how the deep learning method learns certain patterns to detect diseases on medical images. This fact can help the radiologist to understand the deep model and possibly model the pathology assumption in deep network design. There have been several approaches proposed in the literature to visualize and depict feature maps learned by deep models, however, these feature maps are usually not interpretable for radiologists. Hence, potential opportunities exist for designing such methods to characterize the underlying assumption deep models are using and incorporate radiologist feedback inside the network design. 

\section{Conclusion}
In this survey, a detailed discussion regarding the missing modality compensation networks is presented. Our taxonomy divided the literature work into five categories: synthesis models, shared latent space, knowledge distillation networks, mutual information maximization and GANs. For each strategy, a summary of the literature work, network architecture, algorithms, and motivation along with its pros and cons are provided. Furthermore, a detailed discussion of these methods are provided to highlight the most important contribution of each strategy and pointed out the limitation they may face in their design.  Moreover, we summarized the most common benchmarks, evaluation metrics and quantitative performance to provide a clear view of the application for a reader. Finally, our last section provided information regarding the challenges and potential research direction for future work.\\

\textbf{ACKNOWLEDGMENTS}
This work was funded by the German Research Foundation (Deutsche Forschungsgemeinschaft , DFG) under project number 455548460. In addition, it has been funded by the Canada Research Chair in Quantitative Magnetic Resonance Imaging [950-230815], the Canadian Institute of Health Research [CIHR FDN-143263], the Canada Foundation for Innovation [32454, 34824], the Fonds de Recherche du Québec - Santé [322736], the Natural Sciences and Engineering Research Council of Canada [RGPIN-2019-07244], the Canada First Research Excellence Fund (IVADO and TransMedTech), the Courtois NeuroMod project, the Quebec BioImaging Network [5886, 35450], INSPIRED (Spinal Research, UK; Wings for Life, Austria; Craig H. Neilsen Foundation, USA), Mila - Tech Transfer Funding Program.

\bibliographystyle{splncs04}
\bibliography{Refs}

\begin{thebibliography}{100}
\providecommand{\url}[1]{\texttt{#1}}
\providecommand{\urlprefix}{URL }
\providecommand{\doi}[1]{https://doi.org/#1}

\bibitem{asadi2020multi}
Asadi-Aghbolaghi, M., Azad, R., Fathy, M., Escalera, S.: Multi-level context
  gating of embedded collective knowledge for medical image segmentation. arXiv
  preprint arXiv:2003.05056  (2020)

\bibitem{azad2019bi}
Azad, R., Asadi-Aghbolaghi, M., Fathy, M., Escalera, S.: Bi-directional
  convlstm u-net with densley connected convolutions. In: Proceedings of the
  IEEE/CVF International Conference on Computer Vision Workshops. pp.~0--0
  (2019)

\bibitem{azad2020attention}
Azad, R., Asadi-Aghbolaghi, M., Fathy, M., Escalera, S.: Attention deeplabv3+:
  Multi-level context attention mechanism for skin lesion segmentation. In:
  European Conference on Computer Vision. pp. 251--266. Springer (2020)

\bibitem{azad2021deep}
Azad, R., Bozorgpour, A., Asadi-Aghbolaghi, M., Merhof, D., Escalera, S.: Deep
  frequency re-calibration u-net for medical image segmentation. In:
  Proceedings of the IEEE/CVF International Conference on Computer Vision. pp.
  3274--3283 (2021)

\bibitem{azad2021texture}
Azad, R., Fayjie, A.R., Kauffmann, C., Ben~Ayed, I., Pedersoli, M., Dolz, J.:
  On the texture bias for few-shot cnn segmentation. In: Proceedings of the
  IEEE/CVF Winter Conference on Applications of Computer Vision. pp. 2674--2683
  (2021)

\bibitem{azad2021smu}
Azad, R., Khosravi, N., Merhof, D.: Smu-net: Style matching u-net for brain
  tumor segmentation with missing modalities  (2021)

\bibitem{azad2021stacked}
Azad, R., Rouhier, L., Cohen-Adad, J.: Stacked hourglass network with a
  multi-level attention mechanism: Where to look for intervertebral disc
  labeling. In: International Workshop on Machine Learning in Medical Imaging.
  pp. 406--415. Springer (2021)

\bibitem{Baba.2005a}
Baba, Y., Jones, J.: T1 weighted image. In: Radiopaedia.org. Radiopaedia.org
  (2005). \doi{10.53347/rID-5852}

\bibitem{Baba.2005b}
Baba, Y., Niknejad, M.: Fluid attenuated inversion recovery. In:
  Radiopaedia.org. Radiopaedia.org (2005). \doi{10.53347/rID-21760}

\bibitem{web:lang:stats2}
Ballinger, J.R.: Case courtesy of dr j. ray ballinger (2013),
  \url{https://radiopaedia.org/cases/polycystic-ovaries}, last accessed 19
  january 2022

\bibitem{belghazi2018mine}
Belghazi, M.I., Baratin, A., Rajeswar, S., Ozair, S., Bengio, Y., Courville,
  A., Hjelm, R.D.: Mine: mutual information neural estimation. arXiv preprint
  arXiv:1801.04062  (2018)

\bibitem{bessadok2021brain}
Bessadok, A., Mahjoub, M.A., Rekik, I.: Brain graph synthesis by dual
  adversarial domain alignment and target graph prediction from a source graph.
  Medical Image Analysis  \textbf{68},  101902 (2021)

\bibitem{biondetti2021pet}
Biondetti, P., Vangel, M.G., Lahoud, R.M., Furtado, F.S., Rosen, B.R., Groshar,
  D., Canamaque, L.G., Umutlu, L., Zhang, E.W., Mahmood, U., et~al.: Pet/mri
  assessment of lung nodules in primary abdominal malignancies: sensitivity and
  outcome analysis. European Journal of Nuclear Medicine and Molecular Imaging
  \textbf{48}(6),  1976--1986 (2021)

\bibitem{bleker2021single}
Bleker, J., Yakar, D., van Noort, B., Rouw, D., de~Jong, I.J., Dierckx, R.A.,
  Kwee, T.C., Huisman, H.: Single-center versus multi-center biparametric mri
  radiomics approach for clinically significant peripheral zone prostate
  cancer. Insights into imaging  \textbf{12}(1),  1--11 (2021)

\bibitem{bozorgpour2021multi}
Bozorgpour, A., Azad, R., Showkatian, E., Sulaiman, A.: Multi-scale regional
  attention deeplab3+: Multiple myeloma plasma cells segmentation in
  microscopic images. arXiv preprint arXiv:2105.06238  (2021)

\bibitem{brady2017error}
Brady, A.: Error and discrepancy in radiology—inevitable or avoidable?
  insights imaging 8: 171--182 (2017)

\bibitem{brosch2015deep}
Brosch, T., Yoo, Y., Tang, L.Y., Li, D.K., Traboulsee, A., Tam, R.: Deep
  convolutional encoder networks for multiple sclerosis lesion segmentation.
  In: International conference on medical image computing and computer-assisted
  intervention. pp. 3--11. Springer (2015)

\bibitem{brown2011mri}
Brown, M.A., Semelka, R.C.: MRI: basic principles and applications. John Wiley
  \& Sons (2011)

\bibitem{brudfors2019empirical}
Brudfors, M., Ashburner, J., Nachev, P., Balbastre, Y.: Empirical bayesian
  mixture models for medical image translation. In: International Workshop on
  Simulation and Synthesis in Medical Imaging. pp. 1--12. Springer (2019)

\bibitem{bucilua2006model}
Buciluǎ, C., Caruana, R., Niculescu-Mizil, A.: Model compression. In:
  Proceedings of the 12th ACM SIGKDD international conference on Knowledge
  discovery and data mining. pp. 535--541 (2006)

\bibitem{cao2020auto}
Cao, B., Zhang, H., Wang, N., Gao, X., Shen, D.: Auto-gan: self-supervised
  collaborative learning for medical image synthesis. In: Proceedings of the
  AAAI Conference on Artificial Intelligence. vol.~34, pp. 10486--10493 (2020)

\bibitem{carass2017longitudinal}
Carass, A., Roy, S., Jog, A., Cuzzocreo, J.L., Magrath, E., Gherman, A.,
  Button, J., Nguyen, J., Bazin, P.L., Calabresi, P.A., et~al.: Longitudinal
  multiple sclerosis lesion segmentation data resource. Data in brief
  \textbf{12},  346--350 (2017)

\bibitem{chang2020multi}
Chang, Q., Yan, Z., Baskaran, L., Qu, H., Zhang, Y., Zhang, T., Zhang, S.,
  Metaxas, D.N.: Multi-modal asyndgan: Learn from distributed medical image
  data without sharing private information. arXiv preprint arXiv:2012.08604
  (2020)

\bibitem{chen2019robust}
Chen, C., Dou, Q., Jin, Y., Chen, H., Qin, J., Heng, P.A.: Robust multimodal
  brain tumor segmentation via feature disentanglement and gated fusion. In:
  International Conference on Medical Image Computing and Computer-Assisted
  Intervention. pp. 447--456. Springer (2019)

\bibitem{chen2021learning}
Chen, C., Dou, Q., Jin, Y., Liu, Q., Heng, P.A.: Learning with privileged
  multimodal knowledge for unimodal segmentation. IEEE Transactions on Medical
  Imaging  (2021)

\bibitem{chen2020anatomy}
Chen, X., Lian, C., Wang, L., Deng, H., Kuang, T., Fung, S., Gateno, J., Yap,
  P.T., Xia, J.J., Shen, D.: Anatomy-regularized representation learning for
  cross-modality medical image segmentation. IEEE Transactions on Medical
  Imaging  \textbf{40}(1),  274--285 (2020)

\bibitem{cciccek20163d}
{\c{C}}i{\c{c}}ek, {\"O}., Abdulkadir, A., Lienkamp, S.S., Brox, T.,
  Ronneberger, O.: 3d u-net: learning dense volumetric segmentation from sparse
  annotation. In: International conference on medical image computing and
  computer-assisted intervention. pp. 424--432. Springer (2016)

\bibitem{commowick2021multiple}
Commowick, O., Kain, M., Casey, R., Ameli, R., Ferr{\'e}, J.C., Kerbrat, A.,
  Tourdias, T., Cervenansky, F., Camarasu-Pop, S., Glatard, T., et~al.:
  Multiple sclerosis lesions segmentation from multiple experts: The miccai
  2016 challenge dataset. NeuroImage  \textbf{244},  118589 (2021)

\bibitem{conze2021abdominal}
Conze, P.H., Kavur, A.E., Cornec-Le~Gall, E., Gezer, N.S., Le~Meur, Y., Selver,
  M.A., Rousseau, F.: Abdominal multi-organ segmentation with cascaded
  convolutional and adversarial deep networks. Artificial Intelligence in
  Medicine  \textbf{117},  102109 (2021)

\bibitem{dalmaz2021resvit}
Dalmaz, O., Yurt, M., {\c{C}}ukur, T.: Resvit: Residual vision transformers for
  multi-modal medical image synthesis. arXiv preprint arXiv:2106.16031  (2021)

\bibitem{ding2021rfnet}
Ding, Y., Yu, X., Yang, Y.: Rfnet: Region-aware fusion network for incomplete
  multi-modal brain tumor segmentation. In: Proceedings of the IEEE/CVF
  International Conference on Computer Vision. pp. 3975--3984 (2021)

\bibitem{dinsdale2021learning}
Dinsdale, N.K., Bluemke, E., Smith, S.M., Arya, Z., Vidaurre, D., Jenkinson,
  M., Namburete, A.I.: Learning patterns of the ageing brain in mri using deep
  convolutional networks. Neuroimage  \textbf{224},  117401 (2021)

\bibitem{dorent2019hetero}
Dorent, R., Joutard, S., Modat, M., Ourselin, S., Vercauteren, T.: Hetero-modal
  variational encoder-decoder for joint modality completion and segmentation.
  In: International Conference on Medical Image Computing and Computer-Assisted
  Intervention. pp. 74--82. Springer (2019)

\bibitem{fei2021deep}
Fei, Y., Zhan, B., Hong, M., Wu, X., Zhou, J., Wang, Y.: Deep learning-based
  multi-modal computing with feature disentanglement for mri image synthesis.
  Medical Physics  (2021)

\bibitem{feng2021brain}
Feng, C.M., Wang, K., Lu, S., Xu, Y., Li, X.: Brain mri super-resolution using
  coupled-projection residual network. Neurocomputing  \textbf{456},  190--199
  (2021)

\bibitem{feyjie2020semi}
Feyjie, A.R., Azad, R., Pedersoli, M., Kauffman, C., Ayed, I.B., Dolz, J.:
  Semi-supervised few-shot learning for medical image segmentation. arXiv
  preprint arXiv:2003.08462  (2020)

\bibitem{fischer2012introduction}
Fischer, A., Igel, C.: An introduction to restricted boltzmann machines. In:
  Iberoamerican congress on pattern recognition. pp. 14--36. Springer (2012)

\bibitem{geremia2013spatially}
Geremia, E., Menze, B.H., Ayache, N.: Spatially adaptive random forests. In:
  2013 IEEE 10th International Symposium on Biomedical Imaging. pp. 1344--1347.
  IEEE (2013)

\bibitem{giacomello2019transfer}
Giacomello, E., Loiacono, D., Mainardi, L.: Transfer brain mri tumor
  segmentation models across modalities with adversarial networks. arXiv
  preprint arXiv:1910.02717  (2019)

\bibitem{goodfellow2014generative}
Goodfellow, I., Pouget-Abadie, J., Mirza, M., Xu, B., Warde-Farley, D., Ozair,
  S., Courville, A., Bengio, Y.: Generative adversarial nets. Advances in
  neural information processing systems  \textbf{27} (2014)

\bibitem{graves2013body}
Graves, M.J., Mitchell, D.G.: Body mri artifacts in clinical practice: a
  physicist's and radiologist's perspective. Journal of Magnetic Resonance
  Imaging  \textbf{38}(2),  269--287 (2013)

\bibitem{hamghalam2021modality}
Hamghalam, M., Frangi, A.F., Lei, B., Simpson, A.L.: Modality completion via
  gaussian process prior variational autoencoders for multi-modal glioma
  segmentation. In: International Conference on Medical Image Computing and
  Computer-Assisted Intervention. pp. 442--452. Springer (2021)

\bibitem{Haouimi.2005}
Haouimi, A., Jones, J.: T2 weighted image. In: Radiopaedia.org. Radiopaedia.org
  (2005). \doi{10.53347/rID-6345}

\bibitem{havaei2016hemis}
Havaei, M., Guizard, N., Chapados, N., Bengio, Y.: Hemis: Hetero-modal image
  segmentation. In: International Conference on Medical Image Computing and
  Computer-Assisted Intervention. pp. 469--477. Springer (2016)

\bibitem{hinton2015distilling}
Hinton, G., Vinyals, O., Dean, J.: Distilling the knowledge in a neural
  network. arXiv preprint arXiv:1503.02531  (2015)

\bibitem{hofmann2008mri}
Hofmann, M., Steinke, F., Scheel, V., Charpiat, G., Farquhar, J., Aschoff, P.,
  Brady, M., Sch{\"o}lkopf, B., Pichler, B.J.: Mri-based attenuation correction
  for pet/mri: a novel approach combining pattern recognition and atlas
  registration. Journal of nuclear medicine  \textbf{49}(11),  1875--1883
  (2008)

\bibitem{hu2020knowledge}
Hu, M., Maillard, M., Zhang, Y., Ciceri, T., La~Barbera, G., Bloch, I., Gori,
  P.: Knowledge distillation from multi-modal to mono-modal segmentation
  networks. In: International Conference on Medical Image Computing and
  Computer-Assisted Intervention. pp. 772--781. Springer (2020)

\bibitem{huang2019coca}
Huang, P., Li, D., Jiao, Z., Wei, D., Li, G., Wang, Q., Zhang, H., Shen, D.:
  Coca-gan: common-feature-learning-based context-aware generative adversarial
  network for glioma grading. In: International Conference on Medical Image
  Computing and Computer-Assisted Intervention. pp. 155--163. Springer (2019)

\bibitem{isensee2018no}
Isensee, F., Kickingereder, P., Wick, W., Bendszus, M., Maier-Hein, K.H.: No
  new-net. In: International MICCAI Brainlesion Workshop. pp. 234--244.
  Springer (2018)

\bibitem{islam2021glioblastoma}
Islam, M., Wijethilake, N., Ren, H.: Glioblastoma multiforme prognosis: Mri
  missing modality generation, segmentation and radiogenomic survival
  prediction. Computerized Medical Imaging and Graphics p. 101906 (2021)

\bibitem{isola2017image}
Isola, P., Zhu, J.Y., Zhou, T., Efros, A.A.: Image-to-image translation with
  conditional adversarial networks. In: Proceedings of the IEEE conference on
  computer vision and pattern recognition. pp. 1125--1134 (2017)

\bibitem{jack2008alzheimer}
Jack~Jr, C.R., Bernstein, M.A., Fox, N.C., Thompson, P., Alexander, G., Harvey,
  D., Borowski, B., Britson, P.J., L.~Whitwell, J., Ward, C., et~al.: The
  alzheimer's disease neuroimaging initiative (adni): Mri methods. Journal of
  Magnetic Resonance Imaging: An Official Journal of the International Society
  for Magnetic Resonance in Medicine  \textbf{27}(4),  685--691 (2008)

\bibitem{jog2017random}
Jog, A., Carass, A., Roy, S., Pham, D.L., Prince, J.L.: Random forest
  regression for magnetic resonance image synthesis. Medical image analysis
  \textbf{35},  475--488 (2017)

\bibitem{kamnitsas2017efficient}
Kamnitsas, K., Ledig, C., Newcombe, V.F., Simpson, J.P., Kane, A.D., Menon,
  D.K., Rueckert, D., Glocker, B.: Efficient multi-scale 3d cnn with fully
  connected crf for accurate brain lesion segmentation. Medical image analysis
  \textbf{36},  61--78 (2017)

\bibitem{kavur2021chaos}
Kavur, A.E., Gezer, N.S., Bar{\i}{\c{s}}, M., Aslan, S., Conze, P.H., Groza,
  V., Pham, D.D., Chatterjee, S., Ernst, P., {\"O}zkan, S., et~al.: Chaos
  challenge-combined (ct-mr) healthy abdominal organ segmentation. Medical
  Image Analysis  \textbf{69},  101950 (2021)

\bibitem{krupa2015artifacts}
Krupa, K., Bekiesi{\'n}ska-Figatowska, M.: Artifacts in magnetic resonance
  imaging. Polish journal of radiology  \textbf{80}, ~93 (2015)

\bibitem{web:lang:stats5}
Kuijf, H.J.: Miccai-wmh dataset (2017), \url{https://wmh.isi.uu.nl/}, last
  accessed 19 january 2022

\bibitem{lau2019unified}
Lau, K., Adler, J., Sj{\"o}lund, J.: A unified representation network for
  segmentation with missing modalities. arXiv preprint arXiv:1908.06683  (2019)

\bibitem{lee2018efficiency}
Lee, Y.H.: Efficiency improvement in a busy radiology practice: determination
  of musculoskeletal magnetic resonance imaging protocol using deep-learning
  convolutional neural networks. Journal of digital imaging  \textbf{31}(5),
  604--610 (2018)

\bibitem{li2019diamondgan}
Li, H., Paetzold, J.C., Sekuboyina, A., Kofler, F., Zhang, J., Kirschke, J.S.,
  Wiestler, B., Menze, B.: Diamondgan: unified multi-modal generative
  adversarial networks for mri sequences synthesis. In: International
  Conference on Medical Image Computing and Computer-Assisted Intervention. pp.
  795--803. Springer (2019)

\bibitem{maier2017isles}
Maier, O., Menze, B.H., von~der Gablentz, J., H{\"a}ni, L., Heinrich, M.P.,
  Liebrand, M., Winzeck, S., Basit, A., Bentley, P., Chen, L., et~al.: Isles
  2015-a public evaluation benchmark for ischemic stroke lesion segmentation
  from multispectral mri. Medical Image Analysis  \textbf{35},  250--269 (2017)

\bibitem{mehta2018rs}
Mehta, R., Arbel, T.: Rs-net: Regression-segmentation 3d cnn for synthesis of
  full resolution missing brain mri in the presence of tumours. In:
  International Workshop on Simulation and Synthesis in Medical Imaging. pp.
  119--129. Springer (2018)

\bibitem{menze2014multimodal}
Menze, B.H., Jakab, A., Bauer, S., Kalpathy-Cramer, J., Farahani, K., Kirby,
  J., Burren, Y., Porz, N., Slotboom, J., Wiest, R., et~al.: The multimodal
  brain tumor image segmentation benchmark (brats). IEEE transactions on
  medical imaging  \textbf{34}(10),  1993--2024 (2014)

\bibitem{milletari2016v}
Milletari, F., Navab, N., Ahmadi, S.A.: V-net: Fully convolutional neural
  networks for volumetric medical image segmentation. In: 2016 fourth
  international conference on 3D vision (3DV). pp. 565--571. IEEE (2016)

\bibitem{web:lang:stats}
Mudgal, P.: Case courtesy of dr prashant mudgal (2012),
  \url{https://radiopaedia.org/cases/26952/studies/27131}, last accessed 19
  january 2022

\bibitem{myronenko20183d}
Myronenko, A.: 3d mri brain tumor segmentation using autoencoder
  regularization. In: International MICCAI Brainlesion Workshop. pp. 311--320.
  Springer (2018)

\bibitem{orbes2018simultaneous}
Orbes-Arteaga, M., Cardoso, M.J., S{\o}rensen, L., Modat, M., Ourselin, S.,
  Nielsen, M., Pai, A.: Simultaneous synthesis of flair and segmentation of
  white matter hypointensities from t1 mris. arXiv preprint arXiv:1808.06519
  (2018)

\bibitem{ouyang2021representation}
Ouyang, J., Adeli, E., Pohl, K.M., Zhao, Q., Zaharchuk, G.: Representation
  disentanglement for multi-modal brain mri analysis. In: International
  Conference on Information Processing in Medical Imaging. pp. 321--333.
  Springer (2021)

\bibitem{pan2021collaborative}
Pan, Y., Chen, Y., Shen, D., Xia, Y.: Collaborative image synthesis and disease
  diagnosis for classification of neurodegenerative disorders with incomplete
  multi-modal neuroimages. In: International Conference on Medical Image
  Computing and Computer-Assisted Intervention. pp. 480--489. Springer (2021)

\bibitem{park2021prediction}
Park, Y.M., Lim, J.Y., Koh, Y.W., Kim, S.H., Choi, E.C.: Prediction of
  treatment outcome using mri radiomics and machine learning in oropharyngeal
  cancer patients after surgical treatment. Oral Oncology  \textbf{122},
  105559 (2021)

\bibitem{pizzi2021mri}
Pizzi, A.D., Chiarelli, A.M., Chiacchiaretta, P., d’Annibale, M., Croce, P.,
  Rosa, C., Mastrodicasa, D., Trebeschi, S., Lambregts, D.M.J., Caposiena, D.,
  et~al.: Mri-based clinical-radiomics model predicts tumor response before
  treatment in locally advanced rectal cancer. Scientific Reports
  \textbf{11}(1),  1--11 (2021)

\bibitem{article.motion}
Rauf, N., Alam, D., Jamaluddin, M., Samad, B.: Improve image quality of
  transversal relaxation time propeller and flair on magnetic resonance
  imaging. Journal of Physics: Conference Series  \textbf{979},  012079 (03
  2018). \doi{10.1088/1742-6596/979/1/012079}

\bibitem{reyngoudt2021global}
Reyngoudt, H., Marty, B., Boisserie, J.M., Le~Lou{\"e}r, J., Koumako, C.,
  Baudin, P.Y., Wong, B., Stojkovic, T., B{\'e}hin, A., Gidaro, T., et~al.:
  Global versus individual muscle segmentation to assess quantitative mri-based
  fat fraction changes in neuromuscular diseases. European Radiology
  \textbf{31}(6),  4264--4276 (2021)

\bibitem{reza2022contextual}
Reza, A., Moein, H., Yuli, W., Dorit, M.: Contextual attention network:
  Transformer meets u-net. arXiv preprint arXiv:2203.01932  (2022)

\bibitem{sharma2019missing}
Sharma, A., Hamarneh, G.: Missing mri pulse sequence synthesis using
  multi-modal generative adversarial network. IEEE Transactions on Medical
  Imaging  \textbf{39}(4),  1170--1183 (2019)

\bibitem{shen2019brain}
Shen, Y., Gao, M.: Brain tumor segmentation on mri with missing modalities. In:
  International Conference on Information Processing in Medical Imaging. pp.
  417--428. Springer (2019)

\bibitem{simpson2019large}
Simpson, A.L., Antonelli, M., Bakas, S., Bilello, M., Farahani, K.,
  Van~Ginneken, B., Kopp-Schneider, A., Landman, B.A., Litjens, G., Menze, B.,
  et~al.: A large annotated medical image dataset for the development and
  evaluation of segmentation algorithms. arXiv preprint arXiv:1902.09063
  (2019)

\bibitem{souplet2008automatic}
Souplet, J.C., Lebrun, C., Ayache, N., Malandain, G.: An automatic segmentation
  of t2-flair multiple sclerosis lesions. In: MICCAI-Multiple sclerosis lesion
  segmentation challenge workshop (2008)

\bibitem{styner20083d}
Styner, M., Lee, J., Chin, B., Chin, M., Commowick, O., Tran, H.,
  Markovic-Plese, S., Jewells, V., Warfield, S.: 3d segmentation in the clinic:
  A grand challenge ii: Ms lesion segmentation. Midas Journal  \textbf{2008},
  ~1--6 (2008)

\bibitem{sylvain2020cross}
Sylvain, T., Dutil, F., Berthier, T., Di~Jorio, L., Luck, M., Hjelm, D.,
  Bengio, Y.: Cross-modal information maximization for medical imaging: Cmim.
  arXiv preprint arXiv:2010.10593  (2020)

\bibitem{tanner2012fluid}
Tanner, M., Gambarota, G., Kober, T., Krueger, G., Erritzoe, D., Marques, J.P.,
  Newbould, R.: Fluid and white matter suppression with the mp2rage sequence.
  Journal of Magnetic Resonance Imaging  \textbf{35}(5),  1063--1070 (2012)

\bibitem{tillin2012southall}
Tillin, T., Forouhi, N.G., McKeigue, P.M., Chaturvedi, N.: Southall and brent
  revisited: Cohort profile of sabre, a uk population-based comparison of
  cardiovascular disease and diabetes in people of european, indian asian and
  african caribbean origins. International journal of epidemiology
  \textbf{41}(1),  33--42 (2012)

\bibitem{vadacchino2021had}
Vadacchino, S., Mehta, R., Sepahvand, N.M., Nichyporuk, B., Clark, J.J., Arbel,
  T.: Had-net: A hierarchical adversarial knowledge distillation network for
  improved enhanced tumour segmentation without post-contrast images. arXiv
  preprint arXiv:2103.16617  (2021)

\bibitem{van2015does}
Van~Tulder, G., de~Bruijne, M.: Why does synthesized data improve
  multi-sequence classification? In: International Conference on Medical Image
  Computing and Computer-Assisted Intervention. pp. 531--538. Springer (2015)

\bibitem{varsavsky2018pimms}
Varsavsky, T., Eaton-Rosen, Z., Sudre, C.H., Nachev, P., Cardoso, M.J.: Pimms:
  permutation invariant multi-modal segmentation. In: Deep Learning in Medical
  Image Analysis and Multimodal Learning for Clinical Decision Support, pp.
  201--209. Springer (2018)

\bibitem{wang2020multimodal}
Wang, Q., Zhan, L., Thompson, P., Zhou, J.: Multimodal learning with incomplete
  modalities by knowledge distillation. In: Proceedings of the 26th ACM SIGKDD
  International Conference on Knowledge Discovery \& Data Mining. pp.
  1828--1838 (2020)

\bibitem{wang2021acn}
Wang, Y., Zhang, Y., Liu, Y., Lin, Z., Tian, J., Zhong, C., Shi, Z., Fan, J.,
  He, Z.: Acn: Adversarial co-training network for brain tumor segmentation
  with missing modalities. In: International Conference on Medical Image
  Computing and Computer-Assisted Intervention. pp. 410--420. Springer (2021)

\bibitem{webb2017introduction}
Webb, A.G.: Introduction to biomedical imaging. John Wiley \& Sons (2017)

\bibitem{weishaupt2006does}
Weishaupt, D., K{\"o}chli, V.D., Marincek, B., Froehlich, J.M., Nanz, D.,
  Pruessmann, K.P.: How does MRI work?: an introduction to the physics and
  function of magnetic resonance imaging, vol.~2. Springer (2006)

\bibitem{wu2018multimodal}
Wu, M., Goodman, N.: Multimodal generative models for scalable
  weakly-supervised learning. arXiv preprint arXiv:1802.05335  (2018)

\bibitem{yao2021anisamide}
Yao, W., Liu, C., Wang, N., Zhou, H., Chen, H., Qiao, W.: Anisamide-modified
  dual-responsive drug delivery system with mri capacity for cancer targeting
  therapy. Journal of Molecular Liquids  \textbf{340},  116889 (2021)

\bibitem{yao2021mri}
Yao, W., Liu, C., Wang, N., Zhou, H., Chen, H., Qiao, W.: An mri-guided
  targeting dual-responsive drug delivery system for liver cancer therapy.
  Journal of Colloid and Interface Science  \textbf{603},  783--798 (2021)

\bibitem{yu20183d}
Yu, B., Zhou, L., Wang, L., Fripp, J., Bourgeat, P.: 3d cgan based
  cross-modality mr image synthesis for brain tumor segmentation. In: 2018 IEEE
  15th International Symposium on Biomedical Imaging (ISBI 2018). pp. 626--630.
  IEEE (2018)

\bibitem{yu2015utility}
Yu, H., Buch, K., Li, B., O'Brien, M., Soto, J., Jara, H., Anderson, S.W.:
  Utility of texture analysis for quantifying hepatic fibrosis on proton
  density mri. Journal of Magnetic Resonance Imaging  \textbf{42}(5),
  1259--1265 (2015)

\bibitem{yu2021mousegan}
Yu, Z., Zhai, Y., Han, X., Peng, T., Zhang, X.Y.: Mousegan: Gan-based multiple
  mri modalities synthesis and segmentation for mouse brain structures. In:
  International Conference on Medical Image Computing and Computer-Assisted
  Intervention. pp. 442--450. Springer (2021)

\bibitem{yuan2020unified}
Yuan, W., Wei, J., Wang, J., Ma, Q., Tasdizen, T.: Unified generative
  adversarial networks for multimodal segmentation from unpaired 3d medical
  images. Medical Image Analysis  \textbf{64},  101731 (2020)

\bibitem{zhan2021lr}
Zhan, B., Li, D., Wang, Y., Ma, Z., Wu, X., Zhou, J., Zhou, L.: Lr-cgan: Latent
  representation based conditional generative adversarial network for
  multi-modality mri synthesis. Biomedical Signal Processing and Control
  \textbf{66},  102457 (2021)

\bibitem{zhang2021modality}
Zhang, Y., Yang, J., Tian, J., Shi, Z., Zhong, C., Zhang, Y., He, Z.:
  Modality-aware mutual learning for multi-modal medical image segmentation.
  In: International Conference on Medical Image Computing and Computer-Assisted
  Intervention. pp. 589--599. Springer (2021)

\bibitem{zhao2018deep}
Zhao, X., Wu, Y., Song, G., Li, Z., Zhang, Y., Fan, Y.: A deep learning model
  integrating fcnns and crfs for brain tumor segmentation. Medical image
  analysis  \textbf{43},  98--111 (2018)

\bibitem{zhou2018one}
Zhou, C., Ding, C., Lu, Z., Wang, X., Tao, D.: One-pass multi-task
  convolutional neural networks for efficient brain tumor segmentation. In:
  International Conference on Medical Image Computing and Computer-Assisted
  Intervention. pp. 637--645. Springer (2018)

\bibitem{zhou2020hi}
Zhou, T., Fu, H., Chen, G., Shen, J., Shao, L.: Hi-net: hybrid-fusion network
  for multi-modal mr image synthesis. IEEE Transactions on Medical Imaging
  \textbf{39}(9),  2772--2781 (2020)

\bibitem{zhou2021conditional}
Zhou, T., Canu, S., Vera, P., Ruan, S.: Conditional generator and
  multi-sourcecorrelation guided brain tumor segmentation with missing mr
  modalities. arXiv preprint arXiv:2105.13013  (2021)

\bibitem{zhou2021feature}
Zhou, T., Canu, S., Vera, P., Ruan, S.: Feature-enhanced generation and
  multi-modality fusion based deep neural network for brain tumor segmentation
  with missing mr modalities. Neurocomputing  \textbf{466},  102--112 (2021)

\bibitem{zhou2021latent}
Zhou, T., Canu, S., Vera, P., Ruan, S.: Latent correlation representation
  learning for brain tumor segmentation with missing mri modalities. IEEE
  Transactions on Image Processing  \textbf{30},  4263--4274 (2021)

\bibitem{zhu2021brain}
Zhu, Y., Wang, S., Lin, R., Hu, Y., Chen, Q.: Brain tumor segmentation for
  missing modalities by supplementing missing features. In: 2021 IEEE 6th
  International Conference on Cloud Computing and Big Data Analytics (ICCCBDA).
  pp. 652--656. IEEE (2021)

\end{thebibliography}

\end{document}